\documentclass[12pt,reqno]{amsart}

\usepackage{amsmath,amssymb,amsthm,bm,mathrsfs,braket,sasaki,color,graphicx}

\newtheorem{thm}{Theorem}[section]
\newtheorem{prop}[thm]{Proposition}
\newtheorem{lem}[thm]{Lemma}

\theoremstyle{definition}

\theoremstyle{remark}
\newtheorem{rem}[thm]{Remark}

\def\balpha{\boldsymbol{\alpha}}

\def\Frad{\mathcal{F}_\mathrm{rad}}

\def\laplacian{\triangle}

\title[Spectral Analysis of the Dirac Polaron]{Spectral Analysis of the Dirac Polaron}
\author{Itaru Sasaki}
\address{Department of Mathematical Sciences, Shinshu University, Matsumoto 390--8621, Japan \\}
\email{isasaki@shinshu-u.ac.jp}

\thanks{This work was partly supported by Research supported by KAKENHI Y22740087, and was performed through the Program for Dissemination of Tenure-Track System funded by the Ministry of Education and Science, Japan}

\date{\today}


\begin{document}

\begin{abstract}
A system of a Dirac particle interacting with the radiation field is considered.
The Hamiltonian of the system is defined by
$H = \balpha\cdot(\hat\bp-q\bA(\hat\bx))+m\beta + H_f$, 
where $q\in\BR$ is a coupling constant, $\bA(\hat\bx)$ the quantized vector potential and $H_f$ the free photon Hamiltonian.
Since the total momentum is conserved, $H$ is decomposed with respect to the total momentum
with fiber Hamiltonian $H(\bp), (\bp\in\BR^3)$.
Since the self-adjoint operator $H(\bp)$ is bounded from 
below, one can define the lowest energy $E(\bp,m):=\inf\sigma(H(\bp))$.
We prove that $E(\bp,m)$ is an eigenvalue of $H(\bp)$ under the following conditions:
 (i) infrared regularization and (ii) $E(\bp,m)<E(\bp,0)$.
We also discuss the polarization vectors and the angular momentums.
\end{abstract}
\maketitle

\section{Introduction}
We consider a quantum system of a Dirac particle interacting with the radiation field.
An example of a Dirac particle is the free electron.
The Hilbert space for the Dirac particle is 
\begin{align}
 \cH_\mathrm{p} := L^2(\BR_\bx^3;\BC^4), \label{hp}
\end{align}
and the free Hamiltonian for the Dirac particle is the free Dirac operator $\balpha\cdot \hat\bp+m\beta$
acting on $\cH_\mathrm{p}$,
where $\hat\bp=-i\nabla_\bx$ denotes the momentum for the Dirac particle.
The Hilbert space for the radiation field is the Fock space:
\begin{align}
  \Frad := \bigoplus_{n=0}^\infty \Tensor_\mathrm{sym}^n L^2(\BR_\bk^3\times\{1,2\}), \label{frad}
\end{align}
where $\tensor_\mathrm{sym}^n$ means the $n$-fold symmetric tensor product with
$\tensor_\mathrm{sym}^0 L^2(\BR_\bk^3\times\{1,2\}):=\BC$.
The Hilbert space for the total system is defined by
\begin{align}
  \cH := \cH_\mathrm{p} \tensor \Frad.
\end{align}
In this paper, we consider the quantum system described by the Hamiltonian
\begin{equation}
  H := \balpha\cdot(\hat\bp-q\bA(\hat\bx))+m\beta+H_f, \label{hamil}
\end{equation}
where $q\in\BR$ is a coupling constant, $\bA(\hat\bx)$ denotes the quantized magnetic 
vector potential in the Coulomb gauge and $H_f$ denotes the free photon Hamiltonian.
We impose an ultraviolet cutoff in the quantized vector potential.
We call the quantum system defined by \eqref{hamil} the Dirac-Maxwell model.
The Hamiltonian \eqref{hamil}  was introduced and discussed in the early days in quantum theory(e.g., \cite{Heitler:1954}). By an informal perturbation theory, the Klein-Nishina formula
(which gives a differential cross section for the Compton scattering) can be derived from the 
Dirac-Maxwell model\cite{Heitler:1954}.
A mathematical analysis of the Dirac-Maxwell model was initiated by A. Arai\cite{Arai:1999,Arai:2000}.
In the paper \cite{Arai:2003}, A. Arai proved that a non-relativistic limit of the 
Dirac-Maxwell model converges to the Pauli-Fierz model(the non-relativistic QED).
See also \cite{Arai:2006}.


Since the Hamiltonian $H$ is translation invariant, the total momentum of the system conserved, i.e.,
the Hamiltonian of the system strongly commutes with the total momentum operator
\begin{align}
 \bP  := \hat\bp + d\Gamma(\bk),
\end{align}
where $d\Gamma(\bk)$ denotes the momentum operator of the radiation field.
Hence the Hamiltonian can be decomposed as 
\begin{align}
  H &\cong \int_{\BR^3}^\oplus H(\bp) d \bp, \\
  \mathbf{P} &\cong \int_{\BR^3}^\oplus \bp  d \bp,
\end{align}
where the symbol $\cong$ means a unitary equivalence.
In this paper, we mainly study the fiber Hamiltonian $H(\bp)$ which 
describes the dynamics of the relativistic particle dressed in photons with total momentum $\bp$.
We call the quantum system described by $H(\bp)$ the {\it Dirac polaron}.
As shown in \cite{Arai:2000, Arai:1999}, for $\bp\in\BR^3$, $H(\bp)$ has the form
\begin{equation}
  H(\bp) = \balpha\cdot\bp+m\beta +H_f-\balpha\cdot d\Gamma(\bk)-q\balpha\cdot\bA,
\end{equation}
which acts on $\BC^4\tensor\Frad$,
where $\bA$ denotes the quantized vector potential at the origin($=\bA(\bzero)$).
The fourth term $-\balpha\cdot d\Gamma(\bk)$ describes the reaction due to the radiation field,
and the last term $-q\balpha\cdot\bA$ is the electromagnetic interaction.
It should be noted that $-q\balpha\cdot\bA$ is {\it not} $H(\bp)|_{q=0}$-bounded for any 
nonzero $q$, because the reaction term $-\balpha\cdot d\Gamma(\bk)$ 
is comparable to $H_f$ and $-q\balpha\cdot\bA$ is unbounded.
This fact implies that $-q\balpha\cdot\bA$ is not a small perturbation no matter how $q$ is small.
One of the important fact on the Dirac polaron is that 
$H(\bp)$ is bounded from below for all values of all constants: the 
total momentum $\bp$, the mass $m$ and the coupling constant $q$(see \cite{Sasaki:2005}).
Hence, one can define the lowest energy by
\begin{equation}
E(\bp,m):= \inf\sigma(H(\bp))>-\infty,
\end{equation}
where $\sigma(A)$ denotes the spectrum of $A$.
If $H(\bp)$ has an eigenvalue $E$ for $q\neq 0$, we say that an
 dressed particle state exists and the corresponding eigenvector 
is called a dressed particle state.
In Section \ref{sec 4}, we show that a dressed particle state exists 
under suitable conditions including (i) infrared regularization and (ii) the inequality 
\begin{equation}
  E(\bp,m) < E(\bp,0).   \label{bindd}
\end{equation}
The condition \eqref{bindd} plays an binding condition in Theorem \ref{T1}, \ref{T2} and \ref{T3} below.
One can observe that there exist $m^*>0$ such that \eqref{bindd} holds for all $|m|>m^*$.
We expect that $m^*=0$, but we don't have its proof.
In Section \ref{sec 5}, we study the angular momentum and degeneracy of eigenvalues
of the Dirac polaron $H(\bp)$. We will show that the angular momentum of the $\bp$-direction
commutes with $H(\bp)$, and any eigenvalue of $H(\bp)$ has an even
 multiplicity(admit infinity).
Therefore $E(\bp,m)$ is degenerate if it is an eigenvalue of $H(\bp)$.

This paper has three appendices. In Appendix A, we show that all spectral properties
of the Dirac-Maxwell model and the Dirac polarons are independent of the choice of polarization vectors.
Namely, two Hamiltonians which are defined by different polarization vectors
are unitarily equivalent to each other. The discussions in the Appendix A can be applicable for
various QED models(e.g., Pauli-Fierz model).
In Appendix B, we propose a general definition of the angular momentum.
Although the spectral properties of the QED Hamiltonians are independent of the choice of the polarization
vectors, the definition of the angular momentum depends on the polarization vectors.

In Appendix C, we show some properties of the lowest energy $E(\bp)$ which 
is used in proofs of Theorems \ref{T1}-\ref{T3}.

\section{Definitions of the Model}
In this paper, unless confusion arise, we omit the symbol ``$\tensor$'' between two operators, 
for example, we write $A\tensor I$ as $A$ and $I\tensor B$ as $B$, where $I$ denotes the identity operator.
For a closable operator $T$ on $L^2(\BR_\bk^3\times\{1,2\})$, we denote by
$ d\Gamma(T)$ and $\Gamma(T)$ the second quantization 
operators of $T$ (see \cite{Reed-Simon-II}), which acts on $\Frad$.
For $f\in L^2(\BR_\bk^3\times\{1,2\})$, we denote by $a(f)$ and $a(f)^*$
 the annihilation operator and the the creation operator, respectively(see \cite{Reed-Simon-II}),
which are closed operators acting on $\Frad$.
Let $\be^{(\lambda)}:\BR^3\mapsto\BR^3, \lambda=1,2$,
 be polarization vectors:
\begin{equation*}
 \be^{(\lambda)}(\bk)\cdot \be^{(\mu)}(\bk) = \delta_{\lambda,\mu}, \quad
 \be^{(\lambda)}(\bk)\cdot \bk = \bzero, \quad 
 \bk\in \BR^3, \,  \lambda,\mu\in\{1,2\}.
\end{equation*}
We write as 
$\be^{(\lambda)}(\bk)=(e_1^{(\lambda)}(\bk),e_2^{(\lambda)}(\bk),e_3^{(\lambda)}(\bk))$, 
and we suppose that each component $e^{(\lambda)}_j(\bk)$ is 
a Borel measurable function in $\bk$.
For objects $\ba=(a_1,a_2,a_3)$ and $\bb=(b_1,b_2,b_3)$, 
we set $\mathbf{a}\cdot\mathbf{b}:=\sum_{j=1}^3a_jb_j$ if $\sum_{j=1}^3a_jb_j$.
For a linear $F(\cdot)$ we set $F(\mathbf{a}):= (F(a_1),F(a_2),F(a_3))$.
Let $\omega$ be a multiplication by the function
\begin{equation}
  \ome(\bk) = |\bk|.
\end{equation}
We choose a function 
\begin{equation}
\hat\rho \in L^2(\BR_\bk^3)\cap  \dom(\ome^{-1}),     \label{cond of rho}
\end{equation}
where $\dom$ means the operator domain. 
For $j=1,2,3$ and $\bx\in\BR^3$, we set 
\begin{equation*}
 g_j(\bk,\lambda;\mathbf{x}) := 
 |\bk|^{-1/2}\hat\rho(\bk)e^{(\lambda)}_j(\bk) e^{-i\bk\cdot\bx}, 
 \quad  (\bk,\lambda) \in \BR_\bk^3\times\{1,2\}.
\end{equation*}
For each fixed $\bx \in\BR^3$, the function $g_j(\bx)(\cdot) :=g_j(\cdot \, ;\bx) $ is 
a function in $L^2(\BR_\bk^3\times\{1,2\})$.
The quantized magnetic vector potential  at $\bx\in\BR^3$ is defined by
\begin{align*}
&\bA(\bx):=(A_1(\bx)), A_2(\bx), A_3(\bx)), \\
& A_j(\bx) := \frac{1}{\sqrt{2}}
                    \overline{[a(g_j(\bx))+a(g_j(\bx))^*]}, 
 \quad j=1,2,3,
\end{align*}
where, for a closable operator $T$, $\bar{T}$ denotes its closure. 
For each $\bx\in\BR^3$, $A_j(\bx)$ is a self-adjoint operator 
on $\Frad$(see \cite{Reed-Simon-II}).
Since $\be^{(\lambda)}(\bk)$'s are perpendicular to $\bk$, the operators
$\bA(\bx)$ satisfy the Coulomb gauge condition:
\begin{equation}
   \mathrm{div}\, \bA(\bx) =\sum_{j=1}^3 \del_{x_j}A_j(\bx)= 0.
\end{equation}
\begin{rem}
 The function $\hat\rho$ is called an ultraviolet cutoff function. 
An typical example of $\hat\rho$ is the characteristic function of 
the region $\{\bk\in \BR^3| \kappa\leq |\bk|\leq \Lambda\}$, where
$\kappa$ and $\Lambda$ are non-negative constants.
$\Lambda$ is called an ultraviolet cutoff. $\kappa$ is called an infrared
cutoff if it is strictly positive.
\end{rem}

The Hilbert space $\cH$ can be identified as
\begin{equation}
 \cH = L^2(\BR_\bx^3;\BC^4\tensor\Frad) =
 \int_{\BR^3}^\oplus \BC^4\tensor\Frad  d\bx.   \label{identt}
\end{equation}
Under this identification, we define the quantized vector potential in the following way.
Since $g_j(\bx) \in L^2(\BR_\bk^3\times\{1,2\})$ is strongly continuous in $\bx\in\BR^3$,
the map $\bx \mapsto A_j(\bx)$ is a self-adjoint operator valued measurable function.
Then we can define a self-adjoint operator on $\cH$ by
\begin{equation}
 A_j(\hat{\bx}) := \int_{\BR^3}^\oplus A_j(\bx) d \bx.
\end{equation}
Namely, when we identify $\Psi \in D(A_j(\hat{\bx}))$ as 
the $\Frad$-valued square integrable function and the operator, the action of the 
operator $A_j(\hat\bx)$ is given by $(A_j(\hat\bx)\Psi)(\bx) = A_j(\bx)\Psi(\bx)$, $\bx\in\BR^3$.
The operator valued vector
\begin{align}
  \bA(\hat{\bx}) := (A_1(\hat\bx), A_2(\hat\bx), A_3(\hat\bx))
\end{align}
is also called the quantized vector potential.

The free photon Hamiltonian as the second quantization of $\ome$:
\begin{equation}
 H_f :=  d\Gamma(\omega).
\end{equation}
The Dirac-Maxwell Hamiltonian is defined by
\begin{equation}
 H := \balpha\cdot (\hat{\bp}-q\mathbf{A}(\hat{\bx})) + m\beta 
      + H_f,   \label{defhamil}
\end{equation} 
where $\hat{\bp}=-i\nabla_\bx$ and $\nabla_\bx$ is the gradient operator acting in
$\cH_\mathrm{p}$,  $\balpha=(\alpha_1,\alpha_2,\alpha_3)$ and $\beta$ are Dirac matrices
satisfying $\alpha_1,\alpha_2,\alpha_3,\beta\in M_4(\BC)$ and 
\begin{align}
&  \alpha_j \alpha_k + \alpha_k \alpha_j =2 \delta_{jk}, \\
 &  \alpha_j \beta + \beta \alpha_j =0, \\
  & \beta^2 = I_{\BC^4},
\end{align}
 the constant $m\in\BR$ is the rest mass of the Dirac particle, $q\in\BR$ is a coupling constant.
In the right hand side of \eqref{defhamil}, we omit the symbols $\tensor I$ and $I\tensor$,
i.e., the expression \eqref{defhamil} is an abbreviation for 
\[
 H = (\balpha \cdot \hat\bp+m\beta)\tensor I_{\Frad} -q
 \sum_{j=1}^3(\alpha_j\tensor I_{L^2(\BR_\bx^3)})\cdot A_j(\hat{\bx}) + I_{\cH_\mathrm{p}} \tensor H_f.
\]
In this paper, we use the Weyl representation for the Dirac matrices.
Since all representations of the Dirac matrices are unitarily equivalent to each other, this choice
 does not affect the spectral properties of $H$(see \cite[Lemma 2.25]{Thaller:1992}). 

It is easy to see that $H$ is symmetric. 
Although the essential self-adjointness of $H$ was proven in \cite{Arai:1999}, we give a slightly improved result:
\begin{prop}[Essential self-adjointness]{\label{P sa}}
 $\bar{H}$ is a self-adjoint operator and essentially self-adjoint on any core
 for $\sqrt{-\laplacian}+H_f$.
\end{prop}
\begin{proof}[Proof of Proposition \ref{P sa}]
  The proof is a simple application of Nelson's commutator theorem.
Our choice of a comparison operator for Nelson's commutator theorem is 
$\sqrt{-\triangle}+H_f$. 
See \cite{Sasaki:2006} for details.
\end{proof}

\section{Momentum Conservation and Fiber Hamiltonian $H(\bp)$}
The total momentum operator is defined by
\begin{equation}
 \mathbf{P} := \overline{\hat{\bp} + d\Gamma(\bk)}.
\end{equation}
The Hamiltonian $H$ strongly commutes with $\mathbf{P}$ (see \cite{Arai:1999}).
To construct the fiber Hamiltonian, we define a self-adjoint operator
\begin{equation}
 Q := \overline{\bx\cdot d\Gamma(\bk)}.
\end{equation}
Let $U_F$ be the Fourier transform from $L^2(\BR_\bx^3)$
to $L^2(\BR^3_\bp)$.
We set 
\begin{align}
 & U:= (U_F \tensor I_{\BC^4}) \exp(iQ).
\end{align}
Then we can identify $U\cH$ as a constant fiber direct integral
\begin{equation} \label{decom1}
 U \cH \cong \int_{\BR^3}^\oplus \BC^4\tensor \Frad  d\bp.
\end{equation}
For every $\bp\in\BR^3$, we define
\begin{equation}
 H(\bp) := \balpha\cdot\bp + m\beta + H_f 
             - \balpha\cdot d\Gamma(\bk)
             - q \balpha\cdot \mathbf{A},
\end{equation}
which acts on $\BC^4\tensor \Frad$, where $\mathbf{A}:=\mathbf{A}(\bzero)$.
\begin{prop}{\label{P1.2}}
For all $\bp\in\BR^3$, $H(\bp)$ is essentially self-adjoint and 
\begin{align}
 U \bar{H} U^* &= \int_{\BR^3}^\oplus \overline{{H}(\bp)}  d \bp, \\
 U \mathbf{P} U^* &= \int_{\BR^3}^\oplus \bp d\bp.
\end{align}
hold, where $\int^\oplus(\cdots)$ denotes fiber direct integral operator with respect to \eqref{decom1}.
\end{prop}
 \begin{proof}
See \cite{Arai:2000}.
 \end{proof}

\begin{rem}
 Physically $\overline{H(\bp)}$ is the Hamiltonian of the fixed
total momentum $\bp\in\BR^3$.
One can show that the spectral properties of $\overline{H(\bp)}$
is independent of the choice of polarization vectors, because
the Hamiltonians with different polarization vectors are 
unitarily equivalent each other. See Appendix A.
\end{rem}
\begin{rem}
  We call $H(\bp)$ the Dirac polaron Hamiltonian, which was introduced in \cite{Arai:2006}.
It is expected that, as in the model of the H. Fr\"ohlich polaron, the electromagnetic interaction forms a quasiparticle
where the bare Dirac particle is surrounded by the photon clouds.
Such a quasiparticle with momentum $\bp\in\BR^3$ is considered as the ground state of $\overline{H(\bp)}$,
if it exist.
The existence of ground state of $\overline{H(\bp)}$ is the main subject of our paper.
\end{rem}
\begin{rem}
 Note that $\dom(\balpha\cdot d\Gamma(\bk))\subset \dom(H_f)$.
Hence $\dom(H_f)=\dom(H(\bp))$ and $H(\bp)$ is essentially self-adjoint on $\dom(H_f)$.
\end{rem}
One of the most important fact of $\overline{H(\bp)}$ is the semi-boundedness:
\begin{thm}{\label{T bfb}}(\cite{Sasaki:2005})
For any $\bp$, $\overline{H(\bp)}$ is bounded from below.
Moreover $H(\bp)$ is essentially self-adjoint on any core for $H_f$. 
\end{thm}
\begin{proof}
 The first statement was shown in \cite{Sasaki:2005}, where it is assumed the condition $\hat\rho\in \dom(\ome^{1/2})$, but this should not be included in the proof. The reason is the following.
When the lower bound of $H(\bp)$ is computed in \cite{Sasaki:2005},
it is needed to consider the commutator $[d\Gamma(\bk),a(\mathbf{g})]$, ($\mathbf{g}=(g_1(0),g_2(0),g_3(0))$) which is $-a(\bk\cdot\mathbf{g})$ if $\bk \cdot \mathbf{g}$ is in $L^2$, 
otherwise make no sense as the operators in $\Frad$. 
However, the resulting lower bound is the function of $\norm{\ome^{1/2}\mathbf{g}}_{L^2(\BR^3)}$ but $\norm{\ome\mathbf{g}}_{L^2(\BR^3)}$(see \cite[ineq. (24)]{Sasaki:2005}).
Therefore, firstly, we regularize $\hat\rho$ as $\hat\rho_\lambda(\bk) :=\hat\rho(\bk) \chi_{|\bk|\leq \lambda}$,
then we obtain the lower bound of the regularized Hamiltonian $H_\lambda(\bp) \geq C_\ep$.
Since $C_\lambda$ converges as $\lambda\to \infty$ and 
$H_\bp$ converges to $H(\bp)$ on a finite particle subspace,
we get  $H(\bp) \geq \lim_{\ep\to +0}C_\ep>-\infty$.
The second statement follows from the W\"ust's Theorem(\cite{Reed-Simon-II}) and the bound
\begin{align}
 \norm{\balpha\cdot(d\Gamma(\bk)-q\bA)\Psi}^2 \leq \norm{(H_f+E)\Psi}^2, \quad \Psi\in\dom(H_f)
 \label{wuust}
\end{align}
for some $E>0$. The bound \eqref{wuust} was given in \cite{Sasaki:2005}.
\end{proof}
Thus we can define the lowest energy of the Dirac polaron
with total momentum $\bp$ by:
\begin{equation}
 E(\bp,m) := \inf \sigma(\overline{H(\bp)}).
\end{equation}
The energy $E(\bp,m)$ depends on all parameters $(\bp,m,q)\in\BR^3\times\BR\times\BR$.
When $m$ dependence in $E(\bp,m)$ is not important, we write $E(\bp,m)$ as $E(\bp)$.

\section{Existence of a Ground State}{\label{sec 4}}
For a self-adjoint operator bounded below, $T$, we say that $T$ has a ground state if $\inf{\sigma(T)}$ is an eigenvalue of $T$.
In this section, we give criteria for $\overline{H(\bp)}$ to have a ground state.

\begin{thm}{\label{T1}}
Suppose that $\hat\rho$ is spherically symmetric and the bound
\begin{equation}
 \int_{\BR^3} \frac{q^2}{(E(\bp-\bk)-E(\bp)+|\bk|)^2}
 \frac{|\hat\rho(\bk)|^2}{|\bk|} d\bk <1 \label{eq1}
\end{equation}
holds. Assume that  $E(\bp,m)<E(\bp,0)$.
Then the Dirac polaron Hamiltonian $\overline{H(\bp)}$ has a ground state.
\end{thm}
Using the lower bound on $E(\bp-\bk)-E(\bp)+|\bk|$, which is proved in Theorem \ref{T LBOD} below, 
we obtain the following result:
\begin{thm}{\label{T2}}
Assume that $\hat\rho$ be spherically symmetric and that $E(\bp,m)<E(\bp,0)$.
Assume the infrared regular condition $\hat\rho\in \dom(\ome^{-3/2})$.
Then there exists a constant $q_0> 0$ such that for all $q$ with 
$|q|<q_0$, $\overline{H(\bp)}$ has a ground state.
\end{thm}
 \begin{rem}
 Since $E(\bp,m)$ is concave in $m$(Proposition \ref{P1.3}) and $\lim_{m\to\infty} E(\bp,m)=-\infty$,
 there exist $m^* \geq 0$ such that $E(\bp,m)<E(\bp,0)$ for all $|m|>m^*$.
 \end{rem}
A proof of Theorem \ref{T1} is based on the estimates of a photon number bound.
The condition (\ref{eq1}) can be considered as a restriction on the coupling constant $q$. 
There are two ways to remove this restriction.
The first one is the method discovered by C. G\'erard in \cite{Gerard:2000} and 
another one is the photon derivative bound developed in \cite{Griesemer-Lieb-Loss:2001}.
In this paper, we use the photon derivative bound. We need the additional assumptions:
\begin{itemize}
\item[{($\Lambda$)}] (i) $\hat\rho$ is a spherically symmetric function. 
                    (ii) There is an open set $S\subset \BR^3$ such that 
                    $\bar{S}=\supp\hat\rho$ and $\hat\rho$ is continuously differentiable on $S$.
                    (iii) For all $R>0$, the bounded region $S_R:=\{\bk\in S| |\bk|<R\}$
                    has the cone property(see \cite{lieb-loss-analysis} for the definition).
\end{itemize}

The theorem below proves the existence of ground state of the Dirac polaron
for all values of coupling constant $q$:
\begin{thm}{\label{T3}}
Assume the condition $(\Lambda)$.
Moreover we assume that 
\begin{align}
  \hat\rho\in\dom(\ome^{-3/2}), \quad |\bk|^{-5/2}\hat\rho(\bk) \in L^p(S_R), 
  \quad |\bk|^{-3/2}|\nabla\hat\rho(\bk)| \in L^p(S_R),   \label{star}
\end{align}
for all $p\in[1,2)$ and $R>0$. 
Suppose that $E(\bp,m)<E(\bp,0)$. 
Then $\overline{H(\bp)}$ has a ground state.
\end{thm}
\begin{rem}
 The followings are examples
 Let $\chi_{\kappa,\Lambda}(\bk)$ be a characteristic function of the region
$\{\bk\in\BR^3| \kappa<|\bk|<\Lambda\}$. For all $\kappa>0$ and $\Lambda<\infty$,
the cutoff function $\hat\rho=\chi_{\kappa,\Lambda}$ satisfies the conditions ($\Lambda$)
and (\ref{star}).
The function $\hat\rho(\bk)=|\bk|\exp(-\lambda |\bk|)$ $(\lambda>0)$ also satisfies 
the conditions ($\Lambda$) and (\ref{star}).
\end{rem}


\begin{rem}
 It is known that, in non-relativistic QED, the existence of a dressed particle requires
the restriction $|\bp|/ m \leq 1$ (see \cite{Chen:2001}).
On the other hand, Theorems \ref{T1}-\ref{T3} does not require restriction on $|\bp|/m$.
This fact is a crucial difference between relativistic and non-relativistic dynamics. 
This result can be interpreted as follows.
In general, the velocity operator is defined by $i=\sqrt{-1}$ times the commutator of the energy Hamiltonian with the position. Hence, the velocity operators of the non-relativistic particle
and Dirac particle are defined by
\begin{align}
 \hat\bp/m &= i[\hat\bp^2/2m,\bx], \\
 \balpha   &= [\balpha\cdot\hat\bp+m\beta,\bx],
\end{align}
respectively. 
Hence the non-relativistic particle can move faster than the light,
and the particle with velocity $|\bp|/m>1$ makes a shock wave of light and lose their kinetic energy. Therefore such a non-relativistic particle is unstable in the presence of the 
electromagnetic interaction. 
On the other hand, since the speed of the Dirac particle is smaller than 
that of light, $\norm{\balpha}\leq 1$, this kind of catastrophe does not occur,
and the dressed electron state is stable for all $|\bp|$.
\end{rem}

\begin{rem}
It is easy to see that the Hermitian matrix $\balpha\cdot\bp+m\beta$ has two eigenvalues
 $\pm\sqrt{\bp^2+m^2}$, each of which is two-fold degenerate.
Let $u^{(\pm)}_i\in\BC^4$, $i=1,2$ be the corresponding normalized eigenvectors:
\begin{equation*}
  (\balpha\cdot\bp+m\beta)u^{(\pm)}_i = \pm\sqrt{\bp^2+m^2}u^{(\pm)}_i, \quad i=1,2.
\end{equation*}
Let $\Omega:=(1,0,0,\ldots)\in\Frad$ be the vacuum. 
$\Omega$ is the unique eigenvector of both $H_f$ and $ d\Gamma(k_j)$, $j=1,2,3$.
We set $\Phi^{(\pm)}_i:= u^{(\pm)}_i\tensor \Omega$, $j=1,2$.
Clearly,
\begin{equation*}
  H(\bp)|_{q=0}\Phi^{(\pm)}_i = \pm \sqrt{\bp^2+m^2}\Phi^{(\pm)}_i, \quad i=1,2.
\end{equation*}
Thus, in the case $q=0$, $H(\bp)|_{q=0}$ has two eigenvalues $\pm\sqrt{\bp^2+m^2}$.
These eigenvectors $\Phi^{(+)}_i$, $i=1,2$ (resp. $\Phi_i^{(-)}$, $i=1,2$) describe states
 of a freely moving positive(resp. negative) energy particle with momentum $\bp$.
Hence, if photons and the Dirac particle are decoupled, a Dirac particle associated with a positive
eigenvalue exists and the positive eigenvalue is embedded.
We are interested in the fate of these eigenvalues when the interaction is switched on.
As is shown in Fig.1, the lowest energy $E(\bp,m)$ converges to $-\sqrt{\bp^2+m^2}$
as $q\to 0$.
 \begin{figure}
   \centering
\unitlength 0.1in
\begin{picture}( 46.3500, 11.5100)(  2.3300,-15.7100)
%
\special{pn 8}%
\special{pa 1784 624}%
\special{pa 4862 624}%
\special{fp}%
%
\special{pn 8}%
\special{pa 1786 1258}%
\special{pa 4866 1258}%
\special{fp}%
%
\special{pn 13}%
\special{pa 3326 1286}%
\special{pa 3326 1232}%
\special{fp}%
%
\special{pn 13}%
\special{pa 3324 596}%
\special{pa 3324 650}%
\special{fp}%
\put(33.1800,-5.2100){\makebox(0,0){0}}%
\put(33.2600,-11.4800){\makebox(0,0){0}}%
%
\special{pn 4}%
\special{sh 0.800}%
\special{pa 2260 1246}%
\special{pa 4866 1246}%
\special{pa 4866 1278}%
\special{pa 2260 1278}%
\special{pa 2260 1246}%
\special{fp}%
\put(40.3200,-5.1500){\makebox(0,0){\small{$\sqrt{\bp^2+m^2}$}}}%
\put(26.1200,-5.0500){\makebox(0,0){\small{$-\sqrt{\bp^2+m^2}$}}}%
\put(22.6000,-14.6900){\makebox(0,0){$E(\bp,m)$}}%
\put(14.4800,-12.5500){\makebox(0,0){$\sigma(\overline{H(\bp)})$}}%
\put(9.3000,-5.7000){\makebox(0,0)[lt]{$\sigma(H(\bp)|_{q=0})$}}%
%
\special{pn 4}%
\special{sh 0.800}%
\special{pa 2608 606}%
\special{pa 4868 606}%
\special{pa 4868 638}%
\special{pa 2608 638}%
\special{pa 2608 606}%
\special{fp}%
%
\special{pn 8}%
\special{pa 2600 670}%
\special{pa 2258 1200}%
\special{fp}%
\special{sh 1}%
\special{pa 2258 1200}%
\special{pa 2312 1154}%
\special{pa 2288 1154}%
\special{pa 2278 1132}%
\special{pa 2258 1200}%
\special{fp}%
%
\special{pn 8}%
\special{sh 0}%
\special{ar 4132 1532 40 40  0.0000000 6.2831853}%
%
\special{pn 8}%
\special{sh 0.800}%
\special{ar 2608 622 40 40  0.0000000 6.2831853}%
%
\special{pn 8}%
\special{sh 0}%
\special{ar 2252 1256 40 40  0.0000000 6.2831853}%
%
\special{pn 8}%
\special{sh 0.800}%
\special{ar 4032 622 40 40  0.0000000 6.2831853}%
%
\special{pn 8}%
\special{pa 4032 682}%
\special{pa 4132 1482}%
\special{dt 0.030}%
\special{sh 1}%
\special{pa 4132 1482}%
\special{pa 4144 1412}%
\special{pa 4126 1428}%
\special{pa 4104 1418}%
\special{pa 4132 1482}%
\special{fp}%
\put(40.3200,-15.3100){\makebox(0,0){?}}%
%
\special{pn 8}%
\special{pa 4278 1422}%
\special{pa 4278 1286}%
\special{fp}%
\special{sh 1}%
\special{pa 4278 1286}%
\special{pa 4258 1352}%
\special{pa 4278 1338}%
\special{pa 4298 1352}%
\special{pa 4278 1286}%
\special{fp}%
%
\special{pn 8}%
\special{pa 4278 1402}%
\special{pa 4278 1532}%
\special{fp}%
\special{sh 1}%
\special{pa 4278 1532}%
\special{pa 4298 1464}%
\special{pa 4278 1478}%
\special{pa 4258 1464}%
\special{pa 4278 1532}%
\special{fp}%
\put(43.2800,-14.1500){\makebox(0,0){?}}%
\end{picture}%
   \caption{Spectrum of $H(\bp)|_{q=0}$ and $H(\bp)$.}
   \label{fig:2}
 \end{figure}
As is written in textbooks of physics(e.g. \cite{Bjorken-Drell:1964, Heitler:1954}), it is expected that any positive energy
 electron falls down to a negative energy states by a spontaneous emission of photons.
Hence it is expected that the eigenvalue $+\sqrt{\bp^2+m^2}$ is unstable under the perturbation $q\balpha\cdot\bA$.
Theorems \ref{T1}-\ref{T3} ensure that the negative energy dressed electron exists
under some conditions.
But the instability of $\sqrt{\bp^2+m^2}$ has not been proved yet.
\end{rem}

\section{Angular Momentum and Degeneracy of Eigenvalues}{\label{sec 5}}
In this section we show that the angular momentum around $\mathbf{j}$-axis
$(\mathbf{j}\in \BR^3\backslash\{0\})$
of the Dirac polaron is conserved if $\bp$ is parallel to $\mathbf{j}$ 
and $\hat\rho(\bk)$ has axial symmetry around $\mathbf{j}$.
Let $(\overline{H(\bp)},\be)$ be a Dirac polaron model with an arbitrarily given
polarization vectors $\be=(\be^{(1)},\be^{(2)})$.
The total angular momentum around $\mathbf{j}$-axis in the system 
$(\overline{H(\bp)},\be)$ is defined by
\begin{equation*}
  J_{\mathbf{j}}(\be) := S_{\mathbf{j}} + L_{\mathbf{j}}(\be),
\end{equation*}
where $S_{\mathbf{j}}:= \oplus^2(\mathbf{j}\cdot\Vec{\sigma})/2$,
$\Vec{\sigma}=(\sigma_1,\sigma_2,\sigma_3)$ are the Pauli matrices, and 
$L_{\mathbf{j}}(\be)$ is a angular momentum for the radiation field,
which is defined in Appendix B.

\begin{prop}
  The spectrum of $J_{\mathbf{j}}(\be)$ is the set of half-integers:
  \begin{equation*}
    \sigma(J_{\mathbf{j}}(\be)) = \mathbb{Z}_{1/2}:=\{\pm1/2,\pm 3/2,\pm 5/2,\cdots\}.
  \end{equation*}
In particular, $J_{\mathbf{j}}(\be)$ is decomposable as
  \begin{align}
     \BC^4\tensor\Frad &\cong \bigoplus_{z\in\mathbb{Z}_{1/2}}\mathcal{F}(z), \notag \\
    J_{\mathbf{j}}(\be)   &\cong \bigoplus_{z\in\mathbb{Z}_{1/2}} z.     \label{decom}
  \end{align}
\end{prop}
The conclusion in this section is the following:
\begin{thm} 
{\label{T deg}}
Let $\mathbf{j}$ be a unit vector being parallel with $\bp$.
  Assume that $\hat\rho(\bk)=\hat\rho(R\bk),\bk\in\BR^3$, for all $R\in O(3)$ with $R\mathbf{j}=\mathbf{j}$.
Then $\overline{H(\bp)}$ strongly commutes with $J_{\mathbf{j}}(\be)$.
In particular, $\overline{H(\bp)}$ is decomposable as
\begin{equation*}
  \overline{H(\bp)} \cong \bigoplus_{z\in\mathbb{Z}_{1/2}} H(\bp:z),
\end{equation*}
corresponding to the decomposition (\ref{decom}). Moreover, for all $z\in\mathbb{Z}_{1/2}$,
$H(\bp:z)$ is unitarily equivalent to $H(\bp:-z)$, and the multiplicity of any eigenvalue of 
$\overline{H(\bp)}$ is even.
\end{thm}
\begin{rem}
 In the paper \cite{Hiroshima:2007}, F. Hiroshima defines an angular momentum in QED,
which differs from our definition.
\end{rem}

\section{Proof of Theorems \ref{T1} - \ref{T3}}
For a constant $\nu\geq 0$, we define a regularized Hamiltonian to avoid the risk of infrared
divergence:
\begin{equation}
 H_\nu(\bp) := \balpha\cdot\bp+m\beta+ H_f(\nu) -\balpha\cdot d\Gamma(\bk)
             -q\balpha\cdot\mathbf{A},
\end{equation}
where
\begin{equation}
  H_f(\nu) :=  d\Gamma(\ome_\nu), \quad \ome_\nu(\bk)= (1+\nu)|\bk|+\nu.
\end{equation}
Let $N_f:=d\Gamma(1)$ be the photon number operator.
Note that $H_f(\nu)=H_f+\nu(H_f+N_f)$ and $H_0(\bp)=H(\bp)$.
By the Kato-Rellich theorem, one can easily show that, for all $\nu>0$,
$H_\nu(\bp)$ is self-adjoint on $\dom(H_f(\nu))$, and essentially self-adjoint on any core for $H_f(\nu)$.
Since $H_\nu(\bp)\geq H(\bp)$, $H_\nu(\bp)$ is also bounded from below.
We set $\cD := \dom(H_f)\cap \dom(N_f)$.
Then $\cD $ is a common core for $\overline{H_\nu(\bp)}$, $(\nu\geq 0)$.
We set
\begin{equation}
  E_\nu(\bp) := \inf \sigma(\overline{H_\nu(\bp)}).
\end{equation}
For $\nu>0$, the massive Hamiltonian $H_\nu(\bp)$ was studied in \cite{Arai:1999, Arai:2000},
in which A. Arai showed that $H_\nu(\bp)$ has a ground state for all $\nu>0$.
\begin{lem}[Existence of ground state for $\nu>0$]{\label{L mgs}}
 Assume that $\nu>0$. Then
\begin{align}
  \inf\sigma_\mathrm{ess}(H_\nu(\bp)) - E_\nu(\bp) \geq \nu.
\end{align}
In particular, $H_\nu(\bp)$ has a ground state.
\end{lem}
\begin{proof}
  See \cite{Arai:2000}.
\end{proof}

By Lemma \ref{L mgs}, for all $\nu>0$, $H_\nu(\bp)$ has a normalized ground state 
$\Phi_\nu(\bp)\in \dom(H_f(\nu))$. 
In the following, we construct a ground state of $H_0(\bp)$ as suitable limits of $\Phi_\nu(\bp)$.
Since $\Phi_\nu(\bp)$ is normalized, there exists a sequence 
$\{\Phi_{\nu_j}(\bp)\}_{j=1}^\infty$ with $\lim_{j\to\infty}\nu_j = 0$
such that $\{\Phi_{\nu_j}\}_j$ has a weak limit. 
 \begin{lem}{\label{L construct}}
Let $\{\nu_j\}_{j=1}^\infty$ be a sequence such that $\Phi_{\nu_j}$ has a weak limit
$\Phi_0(\bp):= \wlim_{j\to\infty}\Phi_{\nu_j}$.
Assume $\Phi_0\neq 0$.
Then $\Phi_0 \in \dom(\overline{H(\bp)})$ and $\Phi_0$ is a ground state of $\overline{H(\bp)}$.
 \end{lem}
 \begin{proof}
For all $\Psi \in \cD$, one has
 \begin{equation}
 \inner{H(\bp)\Psi}{\Phi_0} 
 = \lim_{j\to\infty}\inner{\Psi}{H(\bp)\Phi_{\nu_j}}
 = \lim_{j\to\infty}\inner{\Psi}{ \{E_{\nu_j}(\bp)-\nu_j(H_f+N_f)\}\Phi_{\nu_j}}.
 \end{equation}
By Proposition \ref{P1.9}, we have $E_{\nu_j}(\bp)\to E_0(\bp)$ as $j\to\infty$.
By assumption (2), we have
\begin{equation}
 \lim_{j\to\infty} \nu_j |\inner{\Psi}{(H_f+N_f)\Phi_{\nu_j}}| 
 \leq
 \lim_{j\to\infty} \nu_j \norm{(H_f+N_f)\Psi} \cdot \norm{\Phi_{\nu_j}} =0.
\end{equation}
Hence $\inner{H(\bp)\Psi}{\Phi_0}=\inner{\Psi}{E(\bp)\Phi_0}$ for all $\Psi\in\cD$.
Since $\cD$ is a core for $\overline{H(\bp)}$, $\Phi_0 \in \dom(\overline{H(\bp)})$ and 
$\overline{H(\bp)}\Phi_0 = E(\bp)\Phi_0$ holds.
 \end{proof}

$H_\nu(\bp)$ and $E_\nu(\bp)$ depend on $\bp, m, \nu$, etc.
When we need to indicate its dependence, we write 
$E_\nu(\bp,m,\cdots)$ and $H_\nu(\bp,m,q,\cdots)$ for $E_\nu(\bp)$ and $H_\nu(\bp)$, 
respectively.

In this section, we use the following identification
\begin{equation*}
\BC^4\tensor \Frad =
\bigoplus_{n=0}^\infty \BC^4\tensor \mathcal{F}^{(n)},
 \quad  \mathcal{F}^{(n)}:= \tensor_\mathrm{s}^n L^2(\BR_\bk^3\times\{1,2\}),
\end{equation*}
and each vector $\Psi^{(n)}\in \BC^4\tensor \mathcal{F}^{(n)}$ is 
identified with a Hilbert space valued function 
$\Psi^{(n)}(\bk,\lambda;\cdot)$ $:\BR_\bk^3\times\{1,2\}\mapsto 
\BC^4\tensor\mathcal{F}^{(n-1)}$.
For all$(\bk,\lambda)\in \BR^3\times\{1,2\}$, we define a map
\begin{align}
& a_\lambda(\bk) : \BC^4 \tensor \Frad \to 
     \prod_{n=0}^\infty \BC^4 \tensor \cF^{(n)} := \{(\Phi^{(n)})_{n=0}^\infty| \Phi^{(n)}\in \BC^4\tensor \cF^{(n)}\} \\
& a_\lambda(\bk)\Psi :=
 (\Psi^{(1)}(\bk,\lambda), \sqrt{2}\Psi^{(2)}(\bk,\lambda;\cdot),
  \ldots, \sqrt{n}\Psi^{(n)}(\bk,\lambda;\cdot),\ldots)
 \in
 \prod_{n=0}^\infty \BC^4\tensor\mathcal{F}^{(n)}.
\end{align}
For almost every $(\bk,\lambda)$, $a_\lambda(\bk)$ is well-defined as a linear map.
The smeared annihilation operator $a(f)$ formally satisfies
\begin{align}
 a(f)\Psi = \sum_{\lambda=1,2}\int_{\BR^3}d\bk f(\bk,\lambda)^* a_\lambda(\bk)\Psi.
\end{align}
It is not necessary to consider that $a_\lambda(\bk)$ is an operator valued distribution.
This definition of $a_\lambda(\bk)$ is useful for our purpose below(Proposition \ref{P3.1}).
In general, $a_\lambda(\bk)\Psi\notin \BC^4\tensor\Frad$, but one can show that
$a_\lambda(\bk)\Psi\in \BC^4\tensor\Frad$ for a class of vectors 
$\Psi\in\BC^4\tensor\Frad$.
Let $\mathrm{w}:\BR^3\to[0,\infty)$ be an almost positive Borel measurable function.
Then, for any $\Psi\in \dom( d\Gamma(\mathrm{w})^{1/2})$ and for almost every 
$(\bk,\lambda)\in\BR^3\times\{1,2\}$, the vector $a_\lambda(\bk)\Psi$ 
is a $\BC^4\tensor\Frad$-valued function.
Because, for any $\Psi \in \dom(d\Gamma(\mathrm{w})^{1/2})$, one has
\begin{equation}
 \norm{ d\Gamma(\mathrm{w})^{1/2}\Psi}^2 = 
 \sum_{n=1}^\infty \sum_{\lambda=1,2}\int_{\BR^3} d\bk
  \mathrm{w}(\bk) n \norm{\Psi^{(n)}(\bk,\lambda;\cdot)}_{\BC^4\tensor\mathcal{F}^{(n-1)}}^2 
 <\infty,    \label{ww1}
\end{equation}
and hence $\sum_{n=1}^\infty n\norm{\Psi^{(n)}(\bk,\lambda;\cdot)}_{\BC^4\tensor\cF^{(n-1)}}^2<\infty$
for almost every $(\bk,\lambda)$.

We set $\mathbf{g}(\bk,\lambda):=\mathbf{g}(\bk,\lambda;0)$.
\begin{prop}{\label{P3.1}}
 Let $\nu>0$. Then
 $a_\lambda(\bk)\Phi_\nu(\bp)\in \dom(H_\nu(\bp))$ and 
\begin{equation}
 a_\lambda(\bk)\Phi_\nu(\bp) = \frac{q}{\sqrt{2}}
 (H_\nu(\bp-\bk)-E_\nu(\bp)+\omega_\nu(\bk))^{-1}
 \balpha\cdot \mathbf{g}(\bk,\lambda)\Phi_\nu(\bp), \quad 
 \label{e p7}
\end{equation}
for almost every $(\bk,\lambda)\in\BR^3\times\{1,2\}$.
\end{prop}
\begin{proof}
  For all $f\in\dom(\omega_\nu)$ and $\Psi\in \cD $, we have
\begin{align*}
 \inner{(H_\nu(\bp)-E_\nu(\bp))\Psi}{a(f)\Phi_\nu(\bp)} 
 = 
 \inner{\Psi}{\Big\{-a(\omega_\nu f)+\balpha\cdot a(\bk f)
              +\tfrac{q}{\sqrt{2}}\inner{f}{\mathbf{g}}\Big\}
              \Phi_\nu(\bp)}.
\end{align*}
Hence
\begin{align*}
 &\sum_{\lambda=1,2}\int_{\BR^3} d\bk f(\bk,\lambda)^*
 \inner{(H_\nu(\bp)-E_\nu(\bp))\Psi}{a_\lambda(\bk)\Phi_\nu(\bp)}
 = \\
 &\sum_{\lambda=1,2}\int_{\BR^3} d\bk f(\bk,\lambda)^*
 \inner{\Psi}{-\omega_\nu(\bk)a_\lambda(\bk)\Phi_\nu(\bp)
 +
 \balpha\cdot\bk a_\lambda(\bk)\Phi_\nu(\bp)
 +
 q\balpha\cdot\mathbf{g}(\bk,\lambda)\Phi_\nu(\bp)}.
\end{align*}
Since $\dom(\omega_\nu)$ is dense in $L^2(\BR_\bk^3\times\{1,2\})$, we have
\begin{align*}
& \inner{ (H_\nu(\bp)-E_\nu(\bp))\Psi }{ a_\lambda(\bk)\Phi_\nu(\bp) }  \\
& =
  \langle \Psi , (-\omega_\nu(\bk)a_\lambda(\bk)
 + \balpha\!\cdot\!\bk \,a_\lambda(\bk)
 +  q\balpha\cdot\mathbf{g}(\bk,\lambda))\Phi_\nu(\bp) \rangle ,
\end{align*}
for almost every $(\bk,\lambda)\in\BR^3\times\{1,2\}$, and all $\Psi\in\cD $.
This means that $a_\lambda(\bk)\Phi_\nu(\bp)\in D(H_\nu(\bp))$ and 
\begin{equation*}
 (H_\nu(\bp)-E_\nu(\bp)+\omega_\nu(\bk)-\balpha\cdot\bk)
 a_\lambda(\bk)\Phi_\nu(\bp)
 = \frac{q}{\sqrt{2}}\balpha\cdot\mathbf{g}(\bk,\lambda)\Phi_\nu(\bp).
\end{equation*}
Hence (\ref{e p7}) follows.
\end{proof}
 \begin{lem}{\label{L63u}}
Suppose that $\hat\rho$ is spherically symmetric and $\hat\rho\in \dom(\ome^{-3/2})$. 
Assume that $E(\bp,m)<E(\bp,0)$. Then
\begin{align}
 \limsup_{\nu\to 0} \norm{N_f^{1/2}\Phi_\nu(\bp)}^2 
& \leq   
  \int_{\BR^3} d\bk \frac{q^2}{(E(\bp-\bk) - E(\bp)+|\bk|)^2} 
    \frac{|\hat\rho(\bk)|^2}{|\bk|} 
  <\infty    \label{lu1}\\
 \limsup_{\nu \to 0} \norm{H_f^{1/2}\Phi_\nu(\bp)}^2   
& \leq   
  \int_{\BR^3} d \bk \frac{q^2}{(E(\bp-\bk) - E(\bp)+|\bk|)^2} 
    |\hat\rho(\bk)|^2
 < \infty.   \label{lu2}
\end{align}
 \end{lem}
\begin{proof}
By Proposition \ref{P3.1} and \eqref{ww1} with $\mathrm{w}=1$, we have
\begin{align*}
 \norm{N_f^{1/2}\Phi_\nu(\bp)}^2 
&\leq \sum_{\lambda=1}^2
  \int_{\BR^3} \frac{q^2}{2}
  \frac{\norm{\balpha\cdot\mathbf{g}(\bk,\lambda)\Phi_\nu(\bp)}^2}
       {(E_\nu(\bp-\bk)-E_\nu(\bp)+|\bk|+\nu)^2}    d \bk \\
&=  \int_{\BR^3} \frac{q^2 }
       {(E_\nu(\bp-\bk)-E_\nu(\bp)+|\bk|+\nu)^2} \frac{|\hat\rho(\bk)|^2}{|\bk|}   d \bk.
\end{align*}
By Theorem \ref{T LBOD} and $\hat\rho\in \dom(\ome^{-3/2})$, the right hand side of \eqref{lu1} is finite.
Hence, by Proposition \ref{P1.9} and the Lebesgue convergence theorem, one has \eqref{lu1}.
The proof of \eqref{lu2} is similar. The only thing we have to do is setting $\mathrm{w}(\bk)=\ome(\bk)$.
\end{proof}
\begin{proof}[Proof of Theorem \ref{T1}]
By Proposition \ref{P1.4} , we have
\begin{equation*}
  0\leq E(\bp-\bk) - E(\bp) +|\bk| \leq 2|\bk|.
\end{equation*}
Hence, by \eqref{eq1},
\begin{equation*}
  \frac{q^2}{4}\int_{\BR^3} \frac{|\hat\rho(\bk)|^2}{|\bk|^3}  d\bk
  \leq \int_{\BR^3} \frac{q^2}{(E(\bp-\bk)-E(\bp)+|\bk|)^2} 
       \frac{|\hat\rho(\bk)|^2}{|\bk|}  d\bk  < 1,
\end{equation*}
which implies $\hat\rho\in\dom(\ome^{-3/2})$. 
Hence \eqref{lu1} and \eqref{lu2} holds.

Since $\Phi_\nu(\bp)$ is a unit vector, there exists a subsequence $\nu_j$ such that 
$\nu_j \to 0$ as $j\to\infty$ and $\Phi_0(\bp):= \wlim_{j\to\infty} \Phi_{\nu_j}(\bp)$ exists.
Then, by \eqref{lu1} and \eqref{lu2}, we have
\begin{align*}
 \lim_{j\to\infty}\norm{N_f^{1/2}\Phi_{\nu_j}}<1,  \qquad 
 \lim_{j\to\infty}\norm{H_f^{1/2}\Phi_{\nu_j}}< \infty,
\end{align*}
which implies that $\Phi_0(\bp)\in \dom(N_f^{1/2})\cap \dom(H_f^{1/2})$.
Hence $\Phi_0(\bp)\in Q(\overline{H(\bp)})$, where $Q$ denotes the form domain.
For any $\varphi\in \dom(H(\bp))$, we have
\begin{align*}
 \inner{(H(\bp)-E(\bp))\varphi}{\Phi_0(\bp)}
& =
 \lim_{j\to\infty} \inner{(H(\bp)-E(\bp))\varphi}{\Phi_{\nu_j}(\bp)} \\
& =
 \lim_{j\to\infty} \inner{\varphi}{(E_{\nu_j}(\bp)-E(\bp)-\nu_j(H_f+N_f))\Phi_{\nu_j}(\bp)} \\
& = 0.
\end{align*}
Thus $\Phi_0(\bp) \in \dom(\overline{H(\bp)})$ and $(\overline{H(\bp)}-E(\bp))\Phi_0(\bp)=0$.
Therefore, if $\Phi_0(\bp)\neq 0$, then $\Phi_0(\bp)$ is a ground state of $\overline{H(\bp)}$.
Since $\BC^4$ is a finite dimensional space, the vacuum component
$\Phi_{\nu_j}(\bp)^{(0)}$ strongly converges to $ \Phi_0(\bp)^{(0)}$.
Hence
\begin{equation}
 \norm{\Phi_0(\bp)}^2   \geq \norm{\Phi_0(\bp)^{(0)}}^2
 =
 \lim_{j\to\infty}\norm{\Phi_{\nu_j}(\bp)^{(0)}}^2
 = \lim_{j\to\infty}\inner{\Phi_{\nu_j}(\bp)}{P_\Omega\Phi_{\nu_j}(\bp)},  \label{vaclb}
\end{equation}
where $P_\Omega$ is the orthogonal projection on the vacuum 
$(1,0,0,\ldots)\in\Frad$.
Thus, using \eqref{vaclb} and $N_f\geq 1-P_\Omega$, we have
\begin{equation*}
 \norm{\Phi_0(\bp)}^2  \geq 
 1- \lim_{j\to\infty}\norm{N_f^{1/2}\Phi_{\nu_j}(\bp)}^2 >0.
\end{equation*}
This means that $\Phi_0(\bp)\neq 0$ and $\Phi_0(\bp)$ 
is a ground state of $\overline{H(\bp)}$.
\end{proof}

\begin{proof}[Proof of Theorem \ref{T2}]
Theorem \ref{T2} is immediately derived from Theorem \ref{T1} and Theorem \ref{T LBOD}.
\end{proof}

Next, we prepare some lemmata for the proof of Theorem \ref{T3}.
For a Hilbert space $\mathcal{K}$, we denote by $\mathsf{B}(\mathcal{K})$
 the set of all bounded operators on $\mathcal{K}$.
The next lemma is followed by the second resolvent equation.
\begin{lem}{\label{L3.2}}
Let $\nu>0$. For each $\mathbf{j}\in\BR^3$ with $|\mathbf{j}|=1$,
the operator valued function
$\BR^3\backslash\{\bzero\}: 
\bk \to (H_\nu(\bp-\bk)-E_\nu(\bp)+\ome_\nu(\bk))^{-1}\in
\mathsf{B}(\BC^4\tensor\Frad)$ is differentiable in the sense of operator norm, 
and 
\begin{align*}
 & \del_\mathbf{j}
 (H_\nu(\bp-\bk)-E_\nu(\bp)+\ome_\nu(\bk))^{-1} =\\
 &
 (H_\nu(\bp\!-\!\bk)-E_\nu(\bp)+\ome_\nu(\bk))^{-1}
 \left(\! \balpha\!\cdot\!\mathbf{j} - (1+\nu)\frac{\bk\!\cdot\!\mathbf{j}}{|\bk|}\right)
 (H_\nu(\bp-\bk)-E_\nu(\bp)+\ome_\nu(\bk))^{-1},
\end{align*}
where $\del_\mathbf{j}$ means the $\mathbf{j}$-direction derivative.
\end{lem}
We fix the following polarization vectors in the rest of this section:
\begin{equation}
 \be^{(1)}(\bk) = 
 \frac{(k_2,-k_1,0)}{\sqrt{k_1^2+k_2^2}},
 \quad
 \be^{(2)}(\bk) := \frac{\bk}{|\bk|}\wedge \be^{(1)}(\bk).
 \label{pol vec}
\end{equation}
Now, remember the definition of the set $S$ (which is defined in condition ($\Lambda$)).
We set $\mathsf{X} := S\backslash\{\bk\in\BR^3|k_1=k_2=0\}$,
$\mathsf{X}_R := S_R \cap \mathsf{X}$.
By Lemma \ref{L3.2} and \eqref{pol vec}, we obtain the following result:
\begin{lem}{\label{L3.3}}
Assume the same assumptions as in Theorem \ref{T3}. Then
$a_\lambda(\bk)\Phi_\nu(\bp)$ is strongly 
continuously differentiable in $\mathsf{X}$ and
\begin{align*}
& \del_ja_\lambda(\bk)\Phi_\nu(\bp) \\
&=
 \frac{q}{\sqrt{2}}
 (H_\nu(\bp-\bk)-E_\nu(\bp)+\ome_\nu(\bk))^{-1}
 \left(\alpha_j - (1+\nu)\frac{k_j}{|\bk|}\right) \\
&\quad \times(H_\nu(\bp-\bk)-E_\nu(\bp)+\ome_\nu(\bk))^{-1}
 \balpha\cdot \mathbf{g}(\bk,\lambda)
 \Phi_\nu(\bp) \\
&\quad +
 \frac{q}{\sqrt{2}}
 (H_\nu(\bp-\bk)-E_\nu(\bp)+\ome_\nu(\bk))^{-1}
 \balpha\cdot (\del_j\mathbf{g}(\bk,\lambda))\Phi_\nu(\bp),
\end{align*}
where $\del_j$ denotes the strong derivative in $k_j,(j=1,2,3)$.
\end{lem}
We set 
\begin{equation*}
 \Psi_j(\bk,\lambda) = (\Psi_j^{(n)}(\bk,\lambda;\cdot))_{n=0}^\infty
:=\del_j a_\lambda(\bk)\Phi_\nu(\bp).
\end{equation*}

\begin{lem}{\label{Lemma}}
 Assume the same assumptions as in Theorem \ref{T3}. Then
\begin{align*}
 \del_j\Phi_\nu^{(n)}(\bp)(\bk,\lambda; X; k_2,\ldots,k_n)
 =
 \frac{1}{\sqrt{n}} \Psi_j^{(n-1)}(\bk,\lambda; X; k_2,\ldots,k_n),
 \quad 
 k_\ell =(\bk_\ell,\lambda_\ell), \,
\end{align*}
for all $X\in \{1,2,3,4\}$, $\bk,\bk_\ell\in\mathsf{X}$, $n\in\BN$, $\lambda,\lambda_\ell=1,2$ and
$j=1,2,3$,
where $\del_j$ is the distributional derivative in $k_j$.
\end{lem}
Note that $\del_j$ in the left hand side is a distributional derivative and that in $\Psi_j$ is a strong derivative.
\begin{proof}
 In this proof, for simplicity, we do not indicate $X, \lambda,\lambda_\ell$ and $\bp$.
The operator $\delta_h$ is defined by $\delta_h f(\bk):=f(\bk+h\mathbf{j})-f(\bk)$ for all functions $f(\bk)$.
Let $\psi(\bk,\bk_2,\ldots,\bk_n)\in C_0^\infty(\mathsf{X}^{n+1})$ be arbitrarily.
Clearly, $(\del_j\psi)(\bk,K) = \lim_{h\to 0}h^{-1}(\psi(\bk+h\mathbf{j},K)-\psi(\bk,K))$ uniformly,
where $K=(\bk_2,\ldots,\bk_n)$ and $\mathbf{j}$ is the unit vector of $j$-th axis.
By the definition of the distributional derivative, we have
\begin{align*}
 \int_{\BR^{3n}}d\bk dK \psi(\bk,K) \del_j\Phi_\nu^{(n)}(\bk,K)  
& = -\int_{\BR^{3n}}d\bk dK (\del_j\psi)(\bk,K) \Phi_\nu^{(n)}(\bk,K)  \\
& = -\lim_{h\to 0} \int_{\BR^{3n}}d\bk dK \frac{1}{-h}(\delta_{-h}\psi)(\bk,K) \Phi_\nu^{(n)}(\bk,K)  \\
& =
 \lim_{h\to 0} \int_{\BR^{3n}} d\bk d K \psi(\bk,K)
 \frac{1}{h} (\delta_h \Phi_\nu^{(n)})(\bk,K).
\end{align*}
By Schwarz' inequality, we have
\begin{align}
&\left|
 \int_{\BR^3} d\bk 
 \left[
 \int_{\BR^{3(n-1)}}  \!\!\! d K \psi(\bk,K)
 \left\{
 \frac{1}{h} [\Phi_\nu^{(n)}(\bk+h\mathbf{j},K) -\Phi_\nu^{(n)}(\bk,K)]
 - \frac{1}{\sqrt{n}} \Psi^{(n-1)}(\bk,K)
 \right\}
 \right]
 \right| \notag \\
&\leq \int_{\BR^3}  d\bk \norm{\psi(\bk,\cdot)}_{L^2(\BR^{3(n-1)})}
 \Bignorm{\frac{\delta_{h}}{h}\Phi_\nu^{(n)}(\bk,\cdot)
          - \frac{1}{\sqrt{n}}\Psi^{(n-1)}(\bk,\cdot)}_{L^2(\BR^{3(n-1)})}.
 \label{eq34}
\end{align}
Note that, for all $\bk\in X$, $h^{-1}\delta_h\Phi_\nu^{(n)}(\bk,\cdot)$ strongly converges to
$\frac{1}{\sqrt{n}}\Psi^{(n-1)}(\bk,\cdot)$ in $L^2(\mathsf{X}^{3(n-1)})$
by Lemma \ref{L3.3}.
Moreover, by Lemma \ref{L3.3} and the assumption that $\hat\rho$ is continuously differentiable,
the function $\bk\to\Psi^{(n-1)}(\bk,\cdot)$ is strongly continuous in $\mathsf{X}$.
Set $D$ be the closure of $\{\bk\in\BR^3 | \norm{\psi(\bk,\cdot)}_{L^2(\BR^{3(n-1)})}\neq 0\}$.
Note that $D\subset X$ is a compact set and $d:= \mathrm{dist}(D, X^\mathrm{c})>0$.

For every $\bk\in D$ and $h$ with $|h|<d$, we have
\begin{equation*}
 \frac{\delta_h}{h}\Phi_\nu^{(n)}(\bk,\cdot) =
 \text{s-}\!\! \int_0^1 \frac{1}{\sqrt{n}} 
 \Psi^{(n-1)}(\bk+th\mathbf{j},\cdot) d t,
\end{equation*}
where s-$\int$ means the strong integral in $L^2(X^{3(n-1)})$.
Since $\norm{\Psi^{(n-1)}(\bk,\cdot)}_{L^2(\BR^{3(n-1)})}$ is continuous
in $\bk \in \mathsf{X}$, it is bounded on the compact set $D$.
For any $\bk \in D$ and $|h|<d$, we have
\begin{align*}
&\Bignorm{\frac{\delta_{h}}{|h|}\Phi_\nu^{(n)}(\bk,\cdot)
          - \frac{1}{\sqrt{n}}\Psi^{(n-1)}(\bk,\cdot)}_{L^2(\BR^{3(n-1)})}\\
&\leq 
 \sup_{|t|\leq 1}\frac{1}{\sqrt{n}}
 \norm{\Psi^{(n-1)}(\bk+th\mathbf{j},\cdot)}_{L^2(\BR^{3(n-1)})} 
 + \frac{1}{\sqrt{n}} \norm{\Psi^{(n-1)}(\bk,\cdot)}_{L^2(\BR^{3(n-1)})} \\
&\leq \mathrm{const.},
\end{align*}
where ``const'' means the constant independent of $\bk$ and $h$.
Applying the Lebesgue dominated convergence theorem, we can see the 
 right hand side of (\ref{eq34}) converges to zero as $|h|\to 0$.
\end{proof}
By Lemmas \ref{L3.2}-\ref{L3.3} and direct calculations, we obtain the 
following inequality
\begin{lem}{\label{L3.5}}
Assume the same assumptions as in Theorem \ref{T3}. Then
\begin{align*}
\norm{\del_j a_\lambda(\bk)\Phi_\nu(\bp)} \\
 \leq &
  \frac{|q|}{\sqrt{2}}(2+\nu)(E_\nu(\bp-\bk)-E_\nu(\bp)+\ome_\nu(\bk))^{-2}
  \frac{|\hat\rho(\bk)|}{|\bk|^{1/2}}
 \\
& +\frac{|q|}{\sqrt{2}}(E_\nu(\bp-\bk)-E_\nu(\bp)+\ome_\nu(\bk))^{-1}
 \frac{|\del_j\hat\rho(\bk)|}{|\bk|^{1/2}} \\
& +\frac{|q|}{\sqrt{2}}(E_\nu(\bp-\bk)-E_\nu(\bp)+\ome_\nu(\bk))^{-1}
 \frac{|\hat\rho(\bk)|}{|\bk|^{3/2}}  \\
& +\frac{|q|}{\sqrt{2}}(E_\nu(\bp-\bk)-E_\nu(\bp)+\ome_\nu(\bk))^{-1}
  \frac{|\hat\rho(\bk)|}{|\bk|^{1/2}} |\del_j\be^{(\lambda)}(\bk)|
\end{align*}
for all $\bk\in\mathsf{X}$, $\lambda=1,2$, $j=1,2,3$.
\end{lem}

Our polarization vectors (\ref{pol vec}) satisfy that 
\begin{equation}
  |\del_j \be^{(\lambda)}(\bk)| \leq \frac{2}{\sqrt{k_1^2+k_2^2}}, \quad \text{for }
  \bk\in \BR^3\backslash\{\bk'\in\BR^3|k'_1=k'_2=0\}.   \label{dpolar}
\end{equation}
We set
\begin{align*}
f_\nu^{(1)}(\bk) &:= (E_\nu(\bp-\bk)-E_\nu(\bp)+\ome_\nu(\bk))^{-2}  \frac{|\hat\rho(\bk)|}{|\bk|^{1/2}}\\
f_\nu^{(2)}(\bk)& :=(E_\nu(\bp-\bk)-E_\nu(\bp)+\ome_\nu(\bk))^{-1} 
                 \frac{|\del_j\hat\rho(\bk)|}{|\bk|^{1/2}} \\
f_\nu^{(3)}(\bk)&:=(E_\nu(\bp-\bk)-E_\nu(\bp)+\ome_\nu(\bk))^{-1} \frac{|\hat\rho(\bk)|}{|\bk|^{3/2}}  \\
f_\nu^{(4)}(\bk)&:=(E_\nu(\bp-\bk)-E_\nu(\bp)+\ome_\nu(\bk))^{-1}
  \frac{|\hat\rho(\bk)|}{|\bk|^{1/2}} |\del_j\be^{(\lambda)}(\bk)|.
\end{align*}
\begin{lem}{\label{lpbound}}
  Assume the conditions in Theorem \ref{T3}. Then
  \begin{equation}
    \sup_{0<\nu\leq 1} \norm{f_\nu^{(j)}}_{L^p(S_R)} <\infty, \quad j=1,2,3,4, \quad p\in [1,2).
    \label{lpbound00}
  \end{equation}
\end{lem}
\begin{proof}
First we consider the case $\bp\neq \bzero$. 
Let $b_\nu(\bp)$ be the constant defined in Theorem \ref{T LBOD}.
Since $b_\nu(\bp)$ is continuous in $\nu$ for fixed $\bp$, Theorem \ref{T LBOD} guarantees 
$\sup_{0\leq \nu\leq 1}b_\nu(\bp)= \max_{0\leq \nu\leq 1}b_\nu(\bp) <1$.
By Theorem \ref{T LBOD}, we have
\begin{equation*}
  (E_\nu(\bp-\bk)-E_\nu(\bp)+|\bk|)^{-1} \leq 
   \frac{1}{1-b_\nu(\bp)}\max\Big\{\frac{1}{|\bk|}, \frac{1}{|\bp|} \Big\}
   \leq C \max\Big\{\frac{1}{|\bk|}, \frac{1}{|\bp|} \Big\},
\end{equation*}
where 
\begin{equation*}
  C:= \sup_{0<\nu\leq 1}\frac{1}{1-b_\nu(\bp)}
\end{equation*}
is a finite constant. 
Hence
\begin{equation*}
  f_\nu^{(1)}(\bk) \leq C^2\Big\{ \frac{1}{|\bk|^2}+\frac{1}{|\bp|^2} \Big\}
                \frac{|\hat\rho(\bk)|}{|\bk|^{1/2}}.
\end{equation*}
Since $S_R$ is a bounded region, by the assumption $|\bk|^{-5/2}|\hat\rho(\bk)| \in L^p(S_R)$,
we obtain that 
\begin{equation*}
  \sup_{0<\nu\leq 1} \norm{f_\nu^{(1)}}_{L^p(S_R)} <\infty.
\end{equation*}
Similarly, we obtain that 
\begin{equation*}
  \sup_{0<\nu \leq 1}\norm{f_\nu^{(j)}}_{L^2(S_R)} <\infty, \quad j=2,3.
\end{equation*}
By (\ref{dpolar}), we have
\begin{equation*}
  f_\nu^{(4)}(\bk) 
  \leq 
  C^2\Big\{\frac{1}{|\bk|}+\frac{1}{|\bp|}\Big\}
  \frac{1}{\sqrt{k_1^2+k_2^2}} \frac{|\hat\rho(\bk)|}{|\bk|^{1/2}}.
\end{equation*}
By using the polar coordinate, we have 
\begin{equation*}
  \int_{S_R}f_\nu^{(4)}(\bk)  d\bk \leq 
  2\pi C \int_{[0,\pi)}\sin\theta d\theta \left[\frac{1}{\sin\theta}\right]^p
  \int_{[0,R)} |\bk|^{2-p} \bigg(\frac{|\bk|+|\bp|}{|\bk|\cdot|\bp|}\bigg)^p
  \frac{|\hat\rho(\bk)|^p}{|\bk|} d |\bk|  <\infty.
\end{equation*}
Next we consider the case $\bp=0$.  
By \eqref{e p6} in Proposition \ref{T LBOD}, we have
\begin{align*}
 ( E_\nu(-\bk) - E_\nu(\bzero) + \ome_\nu(\bk) )^{-1} \leq
\begin{cases}
 \frac{P}{a_\nu(P) |\bk|}, \quad \text{if } |\bk| \leq P \\
 a_\nu(P)^{-1}, \quad \text{if}>P,
\end{cases} 
\end{align*}
for any $P>0$. By the similar arguments as above, one can prove \eqref{lpbound00}.
This completes the proof.
\end{proof}

Let $W^{1,p}(\mathcal{X})$ be the Sobolev space on the configuration space $\mathcal{X}$, i.e.,
the set of all $L^p$-functions with its first derivatives are also in $L^p$.
\begin{lem}{\label{L3.6}}
Suppose the same assumptions as in Theorem \ref{T3}. Then
the $n$-th component of the massive ground state satisfies 
$\Phi_\nu^{(n)}\in \oplus^4 W^{1,p}((\mathsf{X}_R\times\{1,2\})^{n})$ for all
$p\in [1,2)$ and all $R>0$, and 
\begin{equation*}
 \sup_{0<\nu<1} 
 \norm{\Phi_\nu^{(n)}(\bp)}_{\oplus^4 W^{1,p}((\mathsf{X}_R\times\{1,2\})^n)} 
 < \infty.
\end{equation*}
\end{lem}
\begin{proof}
By Lemma \ref{Lemma}, we have
\begin{align*}
 (\nabla_\bk a_\lambda(\bk) \Phi_\nu(\bp))^{(n-1)}(X;\bk_1,\lambda_1; \dots; \bk_{n-1},\lambda_{n-1}) \\
 =
 \sqrt{n}\nabla_\bk \Phi_\nu^{(n)}(\bp; X; \bk,\lambda;\bk_1,\lambda_1; \dots; \bk_{n-1},\lambda_{n-1}).
\end{align*}
Using H\"older's inequality and making a change of variables, one has, for all $p<2$,
\begin{align}
& \sum_{X=1}^4\sum_{\lambda_1,\cdots,\lambda_n\in\{1,2\}} \int_{(\mathsf{X}_R)^n} d\bk_1\cdots d\bk_n
 \sum_{i=1}^n \left| \nabla_{\bk_i}\Phi_\nu^{(n)}(\bp; X; \bk_1,\lambda_1;\cdots;\bk_n,\lambda_n)\right|^p \notag \\
& \leq
 C \int_{\mathsf{X}_R} d\bk \norm{\nabla_\bk a_\lambda(\bk)\Phi_\nu(\bp)}^p, \label{inq38}
\end{align}
where $C$ is a constant independent of $\nu$.
By Lemma \ref{L3.5} and Lemma \ref{lpbound},  the right hand side of \eqref{inq38} is finite uniformly in $\nu>0$.
\end{proof}
\begin{proof}[Proof of Theorem \ref{T3}]
As shown in the Proof of Theorem \ref{T1}, 
 there exists a sequence $\{\nu_j\}_{j=1}^\infty$ such that  
$\Phi_0(\bp):= \wlim_{j\to\infty}\Phi_{\nu_j}(\bp)$ exists,
and $\Phi_0(\bp)\in \dom(H_f^{1/2})\cap \dom(N_f^{1/2})$.
Then, $\Phi_0 \in Q(H(\bp))$.
If $\Phi_0(\bp)\neq 0$, then $\Phi_0(\bp)$ is a ground state of $H(\bp)$.
In the following, we show that $\Phi_0(\bp)\neq 0$.

Any vector $\Psi\in \oplus^4 \mathcal{F}^n=\BC^4\tensor\mathcal{F}^n$ 
is a function of 
the particle helicity $X\in\{1,2,3,4\}$, 
the $n$-photon wave number $(\bk_1,\ldots,\bk_n)\in\BR^{3n}$,
and the photon polarization $\lambda_1,\ldots,\lambda_n\in \{1,2\}$.
For simplicity, we set 
\begin{align*}
&\Phi_j^{(n)}(\bk_1,\ldots,\bk_n) :=
 \Phi_{\nu_j}(\bp)^{(n)}(X;\bk_1,\lambda_1;\cdots; \bk_n,\lambda_n), \\
& \Phi_0^{(n)}(\bk_1,\ldots,\bk_n):=
  \Phi_0(\bp)^{(n)}(X;\bk_1,\lambda_1;\ldots;\bk_n,\lambda_n).
\end{align*}
for $X\in\{1,2,3,4\}$ and $\lambda_1,\ldots,\lambda_n\in\{1,2\}$.
Note that $\Phi_j^{(n)},\Phi_0^{(n)}\in L^2(\BR^{3n})$.
We show that $\slim_{j\to\infty}\Phi_j^{(n)}=\Phi_0^{(n)}$ for all
$n\in\BN$, $X\in\{1,2,3,4\}$ and $\lambda_1,\ldots,\lambda_n\in\{1,2\}$.

By Lemma \ref{L3.6} and the Rellich-Kondrashov theorem, it holds that
\begin{align}
  \lim_{j\to\infty}\norm{\Phi_j^{(n)}-\Phi_0^{(n)}}_{L^2(\mathsf{X}_R^n)}=0
\end{align}
for all $R>0$(we refer \cite[page 578]{Griesemer-Lieb-Loss:2001} for details).
We set $\Phi_j:=(\Phi_j^{(n)})_{n=0}^\infty, \Phi_0:=(\Phi_0^{(n)})_{n=0}^\infty \in \oplus^4\Frad$.
Let $\chi_R$ be the characteristic function of the ball 
$\{\bk\in\BR^3||\bk|<R\}$. 
We denote the orthogonal projection onto $\oplus_{i=0}^n\BC^4\tensor\mathcal{F}^j$ by $P_n$.
Then we have
\begin{align*}
 \norm{\Gamma(\chi_R)(\Phi_j-\Phi_0)}^2
&=
 \norm{P_n\Gamma(\chi_R)(\Phi_j-\Phi_0)}^2 + 
 \norm{(1-P_n)\Gamma(\chi_R)(\Phi_j-\Phi_0)}^2 \\
&\leq 
 \norm{P_n\Gamma(\chi_R)(\Phi_j-\Phi_0)}^2 + 
  \frac{1}{n}\norm{N_f^{1/2}\Gamma(\chi_R)(\Phi_j-\Phi_0)}^2.
\end{align*}
Since each component $(\Gamma(\chi_R)\Phi_j)^{(n)}$ converges to
$(\Gamma(\chi_R)\Phi_0)^{(n)}$ strongly as $j\to\infty$, we have
\begin{equation*}
 \limsup_{j\to\infty}\norm{\Gamma(\chi_R)(\Phi_j-\Phi_0)}^2
 \leq 
   \frac{1}{n}
    \limsup_{j\to\infty}\norm{N_f^{1/2}(\Phi_j-\Phi_0)}^2
\end{equation*}
for all $n\in\BN$. 
By Lemma \ref{L63u}, $\limsup_{j\to\infty}\norm{N_f^{1/2}(\Phi_j-\Phi_0)}^2 <\infty$.
Thus we obtain that
\begin{equation}
 \slim_{j\to\infty} \Gamma(\chi_R)\Phi_j = \Gamma(\chi_R)\Phi_0.
 \label{eq 15}
\end{equation}
Therefore for all $R>0$ we have
\begin{align*}
 \norm{\Phi_j-\Phi_0} &= 
 \norm{\Gamma(\chi_R)(\Phi_j-\Phi_0)}
 + \norm{(1-P_0)(\Gamma(\chi_R)-1)(\Phi_j-\Phi_0)}^2\\
 &\leq 
 \norm{\Gamma(\chi_R)(\Phi_j-\Phi_0)}
 + \norm{(1-P_0)(1-\Gamma(\chi_R))H_f^{-1/2}} \cdot
     \norm{H_f^{1/2}(\Phi_j-\Phi_0)} \\
 &\leq
 \norm{\Gamma(\chi_R)(\Phi_j-\Phi_0)}
 + \frac{C}{R^{1/2}}
\end{align*}
where $C$ is a constant independent of $R>0$.
By (\ref{eq 15}), we obtain
\begin{equation*} 
   \slim_{j\to\infty} \Phi_j = \Phi_0,
\end{equation*}
which implies that $\Phi_0$ is a normalized ground state of $\overline{H(\bp)}$.
\end{proof}
%
\section{Proof of Theorem \ref{T deg}}
In this section we assume the assumptions in Theorem \ref{T deg}.
By Appendices A and B, it suffices to prove  Theorem \ref{T deg}
in the case $\be=\bar\be$.
Here $\bar\be$ is the polarization vector defined in (\ref{jsp}).
Note that $\bar\be$ depends on $\mathbf{j}$.
By assumption, there exists a non-negative constant $t$ such
 that $\bp=t\mathbf{j}$.
We choose a matrix $T\in SO(3)$ such that $T^{-1}\bp=(0,0,|\bp|)$ and
 $T^{-1}\mathbf{j}=(0,0,1)$. 
Let $U$ be the unitary operator defined in the proof of
 Proposition \ref{P1.6}. By (\ref{11}), we obtain that 
 \begin{equation*}
   U\overline{H(\bp)}U^*
   = \overline{(|\bp|\alpha_3+m\beta+H_f-\balpha\cdot d\Gamma(\bk)
      -q \balpha\cdot \Phi_{\mathrm{S}}(\Vec{\lambda}))},
 \end{equation*}
where
\begin{equation*}
\Vec{\lambda} = (\lambda_1,\lambda_2,\lambda_3) 
 =  \frac{\hat\rho(T\bk)}{|\bk|^{1/2}}  (T^{-1}\bar\be^{(1)}(T\bk), T^{-1}\bar\be^{(2)}(T\bk))
 \in (L^2(\BR_\bk^3\times\{1,2\}))^3.
\end{equation*}
Since $T\in SO(3)$, we have
\begin{align*}
&  T^{-1}\bar\be^{(1)}(T\bk) = 
 \frac{T^{-1}[(T\bk)\wedge \mathbf{j}]}{|(T\bk)\wedge  \mathbf{j}|}
 = \frac{\bk\wedge  (0,0,1)}{|\bk\wedge (0,0,1)|}, \\
 & T^{-1}\bar\be^{(2)}(T\bk) = \frac{\bk}{|\bk|}\wedge  
  (T^{-1}\bar\be^{(1)}(T\bk)).
\end{align*}
It is easy to see that $\hat\rho(TR'\bk)=\hat\rho(T\bk), \bk\in\BR^3$ for all 
$R'\in O(3)$ such that $R'(0,0,1)=(0,0,1)$.
Since $\mathbf{S}=(i/4)\balpha\wedge \balpha$, we have
\begin{equation*}
  U(\mathbf{j}\cdot\mathbf{S})U^* = 
\frac{i}{4}\mathbf{j}\cdot[(T\balpha)\cdot(T\balpha)]
 = \frac{i}{4}\mathbf{j}\cdot[T (\balpha\wedge \balpha)]
 = \frac{i}{4} (\balpha\wedge \balpha)_3 = S_3.
\end{equation*}
Moreover, one can show that $U(\mathbf{j}\cdot d\Gamma(\Vec{\ell}))U^*= d\Gamma(\ell_3)$.
Therefore,  
\begin{equation*}
  U J_{\mathbf{j}}(\bar\be)U^* = S_3+ d\Gamma(\ell_3),
\end{equation*}
and, hence, we conclude that it is sufficient to prove Theorem \ref{T deg} in the case
\begin{equation}
 \bp=(0,0,|\bp|), \quad \mathbf{j}=(0,0,1).  \label{z}
\end{equation}

\begin{proof}[Proof of Theorem \ref{T deg}]
We assume (\ref{z}) to the end of this proof.  We put 
\begin{equation*}
  \check\be^{(1)}(\bk) := \frac{(k_2,-k_1,0)}{\sqrt{k_1^2+k_2^2}},\quad 
  \check\be^{(2)}(\bk) := \frac{\bk}{|\bk|}\wedge  \check\be^{(1)}(\bk).
\end{equation*}
For a real parameter $\theta\in\BR$, we set
\begin{equation*}
 W:= \exp[i\theta J_{\mathbf{j}}(\check\be)], \quad 
 \Theta :=
  \begin{bmatrix}
    \cos\theta & -\sin\theta & 0 \\
    \sin\theta & \cos\theta  & 0 \\
       0       &   0         & 1 
  \end{bmatrix}.
\end{equation*}
Then we obtain that
\begin{align}
  W \balpha W^* &= \Theta\balpha, \quad W\beta W^*=\beta,  \label{ta1}\\
  W  d\Gamma(\bk)W^* &=\Theta d\Gamma(\bk), \quad 
  W H_f(m) W^* = H_f(m),                                   \label{ta2} \\
  W \mathbf{A} W^* &=  \Theta\mathbf{A}.                  \label{ta3}
\end{align}
Here, to show (\ref{ta3}), we used the specific form of $\check\be$:
\begin{equation*}
  \check\be^{(\lambda)}(\Theta\bk) = \Theta\check\be^{(\lambda)}(\bk),\quad 
  \lambda=1,2.
\end{equation*}
Since $\theta\in\BR$ is arbitrary, (\ref{ta1}),(\ref{ta2}) and (\ref{ta3}) imply
that $\overline{H(\bp)}$ strongly commutes with $J_{\mathbf{j}}(\check\be)$.
Thus, $\overline{H(\bp)}$ is reduced by the projection onto the eigenspace of $J_\mathbf{j}(\check\be)$.
In other words, $\overline{H(\bp)}$ is decomposable as 
\begin{equation*}
  \overline{H(\bp)} \cong \bigoplus_{z\in\mathbb{Z}_{1/2}} H(\bp:z),
\end{equation*}
in the sense of (\ref{decom}). 
We furthermore define unitary operators $\eta,\tau$ and $\Upsilon$ by
\begin{align*}
  (\eta f)(\bk,\lambda) &:=
  \begin{cases}
    -f(k_1,-k_2,k_3,1)   &~~ \text{if}~~ \lambda=1, \\
     f(k_1,-k_2,k_3,2)   &~~ \text{if}~~ \lambda=2, \quad f\in L^2(\BR_\bk^3\times\{1,2\}),
  \end{cases} \\
  \tau &:= \alpha_1\alpha_2\beta, \quad \Upsilon := \tau\cdot\Gamma(\eta).
\end{align*}
It is easy to see that 
\begin{align*}
  \eta \ell_3\eta^* &=-\ell_3, \quad  \tau S_3\tau^* =-S_3, \\
  \eta k_1\eta^*&=k_1,\quad \eta k_2\eta^* =-k_2, \quad \eta k_3\eta^* =k_3, \\
  \tau\alpha_1\tau^*&=\alpha_1, \quad \tau\alpha_2\tau^*=-\alpha_2, \quad 
  \tau\alpha_3\tau^*=\alpha_3, \quad \tau\beta\tau^*=\beta, \\
  \eta \check\be^{(1)}(\bk)\eta^{-1} &= \frac{(k_2,-(-k_1),0)}{\sqrt{k_1^2+k_2^2}}, \quad 
  \eta \check\be^{(2)}(\bk)\eta^{-1} = 
 \frac{(k_1k_3,-k_2k_3,-k_1^2-k_2^2)}{|\bk|\sqrt{k_1^2+k_2^2}}.
\end{align*}
Hence 
\begin{equation*}
  \Upsilon \overline{H(\bp)} \Upsilon^* = \overline{H(\bp)}, \quad 
  \Upsilon J_{\mathbf{j}}\Upsilon^* = -J_{\mathbf{j}}.
\end{equation*}
Let $E(z), z\in\mathbb{Z}_{1/2}$, be the orthogonal projection on $\ker(J_{\mathbf{j}}-z)$.
Note that $\ran(E(z))=\mathcal{F}(z)$.
$E(-z)\Upsilon E(z)$ is a unitary operator from $\ran(E(z))$ to $\ran(E(-z))$
and 
\begin{align*}
  E(-z)\Upsilon E(z) H(\bp:z) E(z)\Upsilon^* E(-z)
 &=E(-z)\Upsilon E(z) \Upsilon^* \overline{H(\bp)} \Upsilon
  E(z) \Upsilon^* E(-z) \\
  &= H(\bp:-z).
\end{align*}
Therefore $H(\bp:z)$ is unitarily equivalent to $H(\bp:-z)$ for 
all $z\in\mathbb{Z}_{1/2}$.
\end{proof}

\appendix
\section{Remarks on the Polarization Vectors}
In this appendix, we show that the quantum electrodynamics is independent
 of the choice of polarization vectors, i.e., 
the Hamiltonians defined by different polarization vectors are 
unitarily equivalent each other. 
We show the equivalence only for the Hamiltonians $H$ and $H(\bp)$,
but one can apply our proof to the Pauli-Fierz model and various QED models.
The proof here is independent of the choice of $\hat\rho$ and $\omega$.

We assume that the polarization vectors $\be^{(1)}(\bk)$, 
$\be^{(2)}(\bk)$ and $\bk$ are a right-handed system;
\begin{equation*}
 \bk \cdot \be^{(1)}(\bk) = 0, \quad 
 \norm{\be^{(1)}(\bk)}_{\BR^3}=1, \quad 
 \be^{(2)}(\bk) = \frac{\bk}{|\bk|} \wedge \be^{(1)}(\bk), \quad 
 \bk \in\BR^3.
\end{equation*}
Next, we take any polarization vectors $\be'^{(1)}$, $\be'^{(2)}$:
\begin{equation*}
 \bk\cdot \be'^{(\lambda)}(\bk) = 0 ,\quad 
 \be'^{(\lambda)}(\bk)\cdot \be'^{(\mu)}(\bk) =  \delta_{\lambda,\mu}, \quad 
 \bk\in\BR^3, \, \lambda,\mu\in\{1,2\}.
\end{equation*}
Let $H'$ and $H'(\bp)$ be the Hamiltonians 
$H$ and $H(\bp)$ with $\be^{(\lambda)}$ replaced by
$\be'^{(\lambda)}$, $\lambda=1,2$, respectively.
\begin{thm}{\label{T AT1}}
 Assume that $H$ is essentially self-adjoint. Then $H'$ is essentially 
self-adjoint and $\bar{H}$ is unitarily equivalent to $\bar{H'}$
by a unitary operator $U(\be\gets\be')$:
\begin{equation*}
  U(\be\gets\be')\bar{H'}U(\be\gets\be')^* =\bar{H}.
\end{equation*}
\end{thm}

\begin{thm}{\label{T AT2}}
 Assume that $H(\bp)$ is essentially self-adjoint. 
Then $H'(\bp)$ is essentially self-adjoint and $\overline{H(\bp)}$ 
is unitarily equivalent to $\overline{H'(\bp)}$:
\begin{equation*}
  U(\be\gets\be')\overline{H'(\bp)}U(\be\gets\be')^* =\overline{H(\bp)}.
\end{equation*}
\end{thm}
\begin{rem}
  The unitary operators $U(\be\gets\be')$ defined below satisfy the chain-rule:
  \begin{align*}
   &U(\be\gets\be') =  U(\be\gets\be'')U(\be''\gets\be') \\
   & U(\be\gets\be')^* = U(\be'\gets\be).
  \end{align*}
\end{rem}

\begin{proof}[Proofs of Theorem \ref{T AT1} and \ref{T AT2}]
 By the definition of polarization vectors, for each $\bk\in\BR^3$
 it holds that $\be'^{(2)}(\bk)=\frac{\bk}{|\bk|}\wedge \be'^{(1)}(\bk)$
or $\be'^{(2)}(\bk)=-\frac{\bk}{|\bk|}\wedge \be'^{(1)}(\bk)$.
Let $O\subset \BR^3$ be the set such that 
$\be'^{(2)}(\bk)=-\frac{\bk}{|\bk|}\wedge \be'^{(1)}(\bk)$, $\bk\in O$, holds.
 We define
\begin{equation*}
 \be''^{(1)}(\bk):=\be'^{(1)}(\bk), \quad
 \be''^{(2)}(\bk):=
 \begin{cases}
   \be'^{(2)}(\bk), \quad \bk\in\BR^3\backslash O, \\
  -\be'^{(2)}(\bk), \quad \bk\in O.
 \end{cases}
\end{equation*}
We define an operator $H''$ by $H$ with 
 $\be^{(\lambda)}$ replaced by $\be''^{(\lambda)}$, $\lambda=1,2$.
Let 
\begin{equation*}
 \mathbf{g}'(\bk,\lambda;\bx) := 
 \frac{\hat\rho(\bk)}{|\bk|^{1/2}} \be'^{(\lambda)}(\bk) e^{-i\bk\cdot\bx}, \quad 
 \mathbf{g}''(\bk,\lambda;\bx) := 
 \frac{\hat\rho(\bk)}{|\bk|^{1/2}} \be''^{(\lambda)}(\bk)e^{-i\bk\cdot\bx}, \quad 
\end{equation*}
and we set
\begin{equation*}
 \mathbf{A}^\sharp(\hat{\bx}) := \!\frac{1}{\sqrt{2}}\!\int_{\BR^3}^\oplus \!
 \overline{[a(\mathbf{g}^\sharp (\cdot,\bx)) +a(\mathbf{g}^\sharp (\cdot,\bx))^*]}  d \bx,
\end{equation*}
where $\sharp$ stands for $'$ and $''$.
Since $(\be''^{(1)}(\bk), \be''^{(2)}(\bk), \bk)$ are
right-handed vectors, i.e.,  $\bk\cdot\be''^{(1)}(\bk)=0$, 
$\be''^{(2)}(\bk)= \frac{\bk}{|\bk|}\wedge \be''^{(1)}(\bk)$,
there exists $\theta(\bk)\in[0,2\pi)$ such that
\begin{equation*}
 \begin{bmatrix}
  \be^{(1)}(\bk) \\
  \be^{(2)}(\bk)
 \end{bmatrix}
 = 
   \begin{bmatrix}
    \cos\theta(\bk)  & -\sin\theta(\bk) \\
     \sin\theta(\bk) & \cos\theta(\bk)
   \end{bmatrix}
   \begin{bmatrix}
    \be''^{(1)}(\bk) \\
    \be''^{(2)}(\bk)
   \end{bmatrix}.
\end{equation*}
We define a unitary operator $u_1$ on $L^2(\BR_\bk^3\times\{1,2\})$ by
\begin{equation*}
 \begin{bmatrix}
  (u_1f)(\bk,1) \\  (u_1f)(\bk,2)
 \end{bmatrix}
 :=
   \begin{bmatrix}
    \cos\theta(\bk)  & -\sin\theta(\bk) \\
     \sin\theta(\bk) & \cos\theta(\bk)
   \end{bmatrix}
   \begin{bmatrix}
    f(\bk,1) \\ f(\bk,2)
   \end{bmatrix}, \quad \bk\in\BR^3.
\end{equation*}
The operator $U(\be\gets\be''):=\Gamma(u_1)$ is a unitary operator on $\Frad$.
It is clear that 
\[ U(\be\gets\be'') d\Gamma(\omega)U(\be\gets\be'')^* =  d\Gamma(\omega). \]
By the equality $u_1\mathbf{g}''(\cdot,\bx)=\mathbf{g}(\cdot,\bx)$,
we have $U(\be\gets\be'')\mathbf{A}''(\hat{\bx})U(\be\gets\be'')^*=\mathbf{A}(\hat{\bx})$.
Therefore we get
\begin{equation*}
 U(\be\gets\be'')\overline{H''}U(\be\gets\be'')^* =
 \overline{U(\be\gets\be'')H''U(\be\gets\be'')^*} =
 \overline{H}.
\end{equation*}
This means that the operator $H''$ is essentially self-adjoint and 
$\overline{H''}$ is unitarily equivalent to $\bar{H}$.
Next we show that $\overline{H''}$ is unitarily equivalent to 
$\overline{H'}$.
Let $u_2$ be a unitary operator on $L^2(\BR_\bk^3\times\{1,2\})$ 
such that 
\begin{equation*}
 (u_2 f)(\bk,\lambda) :=
 \begin{cases}
  -f(\bk,2), \quad\bk\in S, \\
  f(\bk,\lambda), \quad \text{otherwise}. 
 \end{cases}
\end{equation*}
It is easy to see that $u_1g'_j(\cdot,\bx)=g''_j(\cdot,\bx)$,
 $j=1,2,3$. 
Then $U(\be''\gets\be'):=\Gamma(u_2)$ is a unitary transformation on $\Frad$, and 
\begin{equation*}
 U(\be''\gets\be')  d\Gamma(\omega) U(\be''\gets\be')^* =  d\Gamma(\omega).
\end{equation*}
By the definition of $u_2$, the equality
$U(\be''\gets\be')\mathbf{A}'(\hat{\bx})U(\be''\gets\be')^* = \mathbf{A}''(\hat{\bx})$
holds.
Hence we have
\begin{equation*}
 U(\be''\gets\be') \overline{H'} U(\be''\gets\be')^* 
= \overline{U(\be''\gets\be')H'U(\be''\gets\be')^*} =
 \overline{H''},
\end{equation*}
which implies that $H'$ is essentially self-adjoint and 
$\overline{H'}$ is unitarily equivalent to $\overline{H''}$. 
We set
\begin{equation*}
U(\be\gets\be'):= U(\be\gets\be'')U(\be''\gets\be').
\end{equation*}
Then $U(\be\gets\be')\overline{H'}U(\be\gets\be')^*=\bar{H}$.
Therefore Theorem \ref{T AT1} is proved. 
The proof of Theorem \ref{T AT2} is similar to the proof of Theorem \ref{T AT1}.
\end{proof}

\section{Remarks on the Angular Momentum} 
As is shown in Appendix A, spectral properties of QED models are independent
of the choice of polarization vectors.
Hence, in the definition of QED models, usually we do not need to specify the choice of
the polarization vectors.
However, the angular momentum of the electromagnetic field depends on a choice 
of the polarization vectors, since the angular momentum does not commute with
$U(\be\gets\be')$.
Therefore, when we discuss an angular momentum, we take care of specifying
the choice of polarization vectors.
One can find the definition of an angular momentum for the electromagnetic field
in the textbook \cite[Section 13.5]{Spohn:2004}(see also \cite{Hiroshima:2007}).
In this appendix, we propose an alternate definition of angular momentum in the electromagnetic field.

Let $(H,\be)$ be the pair of a Hamiltonian and polarization vectors.

For each unit vector $\mathbf{j}\in\BR^3$,
we can define a specific polarization vectors  $\bar\be=(\bar\be^{(1)},\bar\be^{(2)})$ by 
\begin{equation}
\bar\be^{(1)}(\bk):= \frac{\bk\wedge\mathbf{j}}{|\bk\wedge\mathbf{j}|}, \quad 
 \bar\be{(2)}(\bk):= \frac{\bk}{|\bk|}\wedge \bar\be^{(1)}(\bk).  \label{jsp}
\end{equation}

For a Dirac-Maxwell model $(H,\bar\be)$,
we define the angular momentum around $\mathbf{j}$-axis by
\begin{equation*}
  L_{\mathbf{j}}(\bar\be) :=  d\Gamma(\overline{\mathbf{j}\cdot\Vec{\ell}}),
\end{equation*}
where 
\begin{equation*}
  \Vec{\ell} :=(\ell_1,\ell_2,\ell_3) := i(\nabla_\bk\wedge \bk),
\end{equation*}
is a triplet of self-adjoint operators acting on $L^2(\BR_\bk^3\times\{1,2\})$.

Let $\be=(\be^{(1)},\be^{(2)})$ be any polarization vectors. 
The angular momentum around $\mathbf{j}$-axis in the Dirac-Maxwell model
$(H,\be)$ is defined by
\begin{equation*}
  L_{\mathbf{j}}(\be) := U(\be\gets\bar\be)L_{\mathbf{j}}(\bar\be)U(\be\gets\bar\be)^*,
\end{equation*}
where $U(\bar\be\gets\be)$ is a unitary operator defined in Appendix A.
By the chain-rule of $U(\be\gets\be')$, the angular momentums transformed as
\begin{equation*}
  L_{\mathbf{j}}(\be) = U(\be\gets\be')L_{\mathbf{j}}(\be')U(\be\gets\be')^*,
\end{equation*}
where $\be$ and $\be'$ are arbitrary polarization vectors.


\section{Some Properties of the Lowest Energy}
In Appendix C, we show some properties of $E_\nu(\bp)$ which are used 
in proofs of Theorems \ref{T1}-\ref{T3}.
\begin{prop}[Concavity]{\label{P1.3}}
 $E_\nu(\bp)$ is concave in $(\bp,m,q)\in \BR^3\times\BR\times \BR$.
\end{prop}
\begin{proof}
  See \cite{Arai:2000}.
\end{proof}
\begin{prop}[Continuity]{\label{P1.4}}
 $E_\nu(\bp,m)$ is Lipschitz continuous in $(\bp,m)$, i.e.,
\begin{equation*}
 |E_\nu(\bp,m)-E_\nu(\bp',m')| \leq 
 \sqrt{|\bp-\bp'|^2+|m-m'|^2}, \quad 
 \bp,\bp'\in\BR^3, \, m,m'\in\BR.
\end{equation*}
\end{prop}
\begin{proof}
  See \cite{Arai:2000}
\end{proof}

\begin{prop}[Reflection symmetry in $m$]{\label{P1.5}}
 The Hamiltonian
 $\overline{H_\nu(\bp,m)}$ is unitarily equivalent to 
 $\overline{H_\nu(\bp,-m)}$. 
In particular 
\begin{equation*}
E_\nu(\bp,m)=E_\nu(\bp,-m),  \quad E_\nu(\bp,m) \leq E_\nu(\bp,0).
\end{equation*}
\end{prop}
\begin{proof}
 Let $\gamma_5:=-i\alpha_1\alpha_2\alpha_3$. Then $\gamma_5$ is a unitary 
operator and $\gamma_5\overline{H_\nu(\bp,m)}\gamma_5^* = \overline{H_\nu(\bp,-m)}$.
Therefore $E_\nu(\bp,m)=E_\nu(\bp,-m)$. 
By Proposition \ref{P1.3}, $m\mapsto E_\nu(\bp,m)$ is concave. 
Hence $E_\nu(\bp,0)=E_\nu(\bp,\tfrac{1}{2}m-\tfrac{1}{2}m)\geq E_\nu(\bp,m)$.
\end{proof}

\begin{prop}[Rotation invariance of the total momentum]{\label{P1.6}}
Let $T\in O(3)$ be an orthogonal matrix.
 Assume that $|\hat\rho(\bk)|=|\hat\rho(T\bk)|$ $\, \mathrm{a.e.} \, \bk\in\BR^3$.
Then $\overline{H_\nu(\bp)}$ is unitarily equivalent to $\overline{H_\nu(T\bp)}$.
In particular, $E_\nu(\bp)=E_\nu(T\bp)$ follows.
\end{prop}
\begin{proof}
 For matrix $T\in O(3)$, we define four 4$\times$4 matrices by
\begin{equation*}
 \beta' := \beta, \quad \alpha'_j := \sum_{j=1}^3 T_{j,l}\alpha_l, \quad j=1,2,3,
\end{equation*}
which obeys $\{\alpha'_j,\beta'\}=0$, $\{\alpha'_j,\alpha'_l\}=2 \delta_{j,l}$,
 $j,l=1,2,3$.
 Then there exists a 4$\times$4 unitary matrix
$u_T$ such that (see \cite[Lemma 2.25]{Thaller:1992})
\begin{equation*}
 u_T\alpha_ju_T^{-1} = \sum_{k=1}^3 T_{j,k}\alpha_k, \quad 
 u_T\beta   u_T^{-1} = \beta.
\end{equation*} 
Therefore $u_T\balpha\cdot\bp u_T^{-1} = \sum_{k,l=1}^3T_{l,k}\alpha_k p_l = 
\sum_{k,l=1}^3\alpha_k(T^{-1})_{k,l}p_l = \balpha\cdot(T^{-1}\bp)$.
Similarly, we have
\begin{align*}
 u_T(\balpha\cdot d\Gamma(\bk))u_T^{-1} = 
 \balpha\cdot(T^{-1} d\Gamma(\bk)), \quad 
 u_T \balpha\cdot \mathbf{A} u_T^{-1} = 
 \balpha\cdot (T^{-1}\mathbf{A}) =
 (T\balpha)\cdot \mathbf{A}.
\end{align*} 
We define rotation operator $\hat{T}$ of photon momentum, $\hat{T}$, by 
\begin{equation*}
 (\hat{T}f)(\bk,\lambda) =f(T^{-1}\bk,\lambda), \quad 
 (\bk,\lambda)\in \BR_\bk^3\times\{1,2\}, \quad 
 f\in L^2(\BR_\bk^3\times\{1,2\}).
\end{equation*}
Then for all $f\in \dom(k_j\hat{T})$
\begin{equation*}
\hat{T}^{-1} k_j \hat{T}f(\bk,\lambda) = 
 (k_j\hat{T}f)(T\bk,\lambda) = 
 (T\bk)_j(\hat{T}f)(T\bk,\lambda) = 
 (T\bk)_jf(\bk,\lambda).
\end{equation*}
Hence we obtain the operator equality $\hat{T}^{-1} k_j\hat{T}=(T\bk)_j$, $j=1,2,3$.
Thus
\begin{align*}
 &\Gamma(\hat{T}^{-1})  d\Gamma(k_j) \Gamma(\hat{T})
 =  d\Gamma((T\bk)_j) = (T\cdot d\Gamma(\bk))_j, \\
 &\Gamma(\hat{T}^{-1}) H_f(\nu) \Gamma(\hat{T}) = H_f(\nu) \\
 &\Gamma(\hat{T}^{-1}) A_j \Gamma(\hat{T}) ,
 = \Phi_S(\hat{T}^{-1}g_j), \quad j=1,2,3,
\end{align*}
where $\Phi_S(\cdot)$ is the Segal field operator(see\cite[Page 209]{Reed-Simon-II}) and
$g_j(\cdot):=g_j(\cdot,\bx=\bzero)\in L^2(\BR_\bk^3\times\{1,2\})$.
The operator $U:=u_T\tensor \Gamma(\hat{T}^{-1})$ is a unitary operator on $\BC^4\tensor\Frad$ and
\begin{align}
 U\overline{H_\nu(\bp)}U^{-1} =
 \overline{(\balpha\cdot(T^{-1}\bp) + m\beta + H_f(\nu)
  -\balpha\cdot d\Gamma(\bk) 
  -q (T\balpha)\cdot\Phi_S(\hat{T}^{-1}\mathbf{g}))}. \label{11}
\end{align}
Note that $T$ is a 3$\times$3-matrix and $\hat{T}$ is unitary on $L^2(\BR_\bk^3\times\{1,2\})$.
Since $T\in O(3)$, we have $(T\balpha)\cdot\Phi_S(\hat{T}^{-1}\mathbf{g}) = \balpha\cdot T^{-1}\Phi_S(\hat{T}^{-1}\mathbf{g})$, i.e.,
\begin{equation}
 (T^{-1}\Phi_S(\hat{T}^{-1}\mathbf{g}))_j 
 = 
 \sum_{l=1}^3 (T^{-1})_{j,l} \Phi_S(\hat{T}^{-1}g_l), \quad 
 j=1,2,3.
 \label{12}
\end{equation}
We define functions
\begin{equation*}
 \be'^{(\lambda)}(\bk) = T^{-1}\be^{(\lambda)}(T\bk), \quad 
 (\bk,\lambda)\in \BR^3\times\{1,2\}.
\end{equation*}
Then $\be'^{(1)}$ and $\be'^{(2)}$ are polarization vectors:
$\bk\cdot\be'^{(\lambda)}(\bk)=0,\,\, \be'^{(\lambda)}(\bk)\cdot\be'^{(\mu)}(\bk)= \delta_{\lambda,\mu}$.
Since $|\hat\rho(\bk)|=|\hat\rho(T\bk)|$, there exists a Borel measurable function 
$\bk\mapsto\kappa(\bk)\in\BR$ such that 
$\hat\rho(T\bk)=e^{i\kappa(\bk)}\hat\rho(\bk)$, $\, \mathrm{a.e.} \, \bk\in\BR^3$.
Therefore, we have
\begin{equation}
 \sum_{l=1}^3(T^{-1})_{j,l}g_l(T\bk,\lambda)
 =
 \frac{e^{i\kappa(\bk)}\hat\rho(\bk)}{|\bk|^{1/2}} e_j'^{(\lambda)}(\bk). 
  \label{13}
\end{equation}
Let $H_\nu'(\bp)$ be defined by $H_\nu(\bp)$ with $\be^{(\lambda)}$ replaced by $\be'^{(\lambda)}$. 
By (\ref{11}),(\ref{12}) and (\ref{13}), we have
\begin{equation*}
 U\overline{H_\nu(\bp)}U^* = V\overline{H_\nu'(T^{-1}\bp)}V^*,
\end{equation*}
where $V:=\Gamma(e^{i\kappa(\cdot)})$.
By Theorem \ref{T AT2}, $\overline{H'_\nu(T^{-1}\bp)}$ is unitarily equivalent to
$\overline{H_\nu(T^{-1}\bp)}$. 
Therefore, $\overline{H(\bp)}$ is 
unitarily equivalent to $\overline{H_\nu(T^{-1}\bp)}$.
Since $\bp\in\BR^3$ is arbitrary,
 $\overline{H_\nu(\bp)}$ is unitarily equivalent to $\overline{H_\nu(T\bp)}$,
and  $E_\nu(\bp)=E_\nu(T\bp)$.
\end{proof}

%
If the cutoff function $|\hat\rho(\bk)|$ has the reflection symmetry at the origin,
 the following important inequality holds.
\begin{prop}{\label{P1.7}}
 Assume that $|\hat\rho(\bk)|=|\hat\rho(-\bk)|$ for almost every $\bk\in\BR^3$.
Then the inequality 
\begin{equation*}
E_\nu(\bp) \leq  E_\nu(\bzero) , \quad \bp\in\BR^3\backslash\{\bzero\}
\end{equation*}
holds.
\end{prop}
\begin{proof}
By the assumption $\hat\rho(\bk)=\hat\rho(-\bk)$ $\, \mathrm{a.e.} \,\bk\in\BR^3$ and 
Proposition \ref{P1.6}, we have
$E_\nu(\bp)=E_\nu(-\bp)$, $\bp\in\BR^3$.
Using the concavity of $E_\nu(\bp)$ with respect to $\bp$. 
we obtain
\begin{equation*}
 E_\nu(\bzero) = E_\nu(\tfrac{1}{2}\bp-\tfrac{1}{2}\bp) \geq
          \frac{1}{2}E_\nu(\bp) + \frac{1}{2}E_\nu(-\bp) 
        = E_\nu(\bp)
\end{equation*}
for all $\bp\in\BR^3$.
\end{proof}

Assuming that $H_\nu(\bzero)$ has a ground state,
we can obtain the following strict inverse energy inequality:
\begin{prop}{\label{P1.75}}
Assume that $|\hat\rho(\bk)|=|\hat\rho(-\bk)|$ $\, \mathrm{a.e.} \, \bk\in\BR^3$.
If $\overline{H_\nu(\bzero)}$ has a ground state, then
\begin{equation*}
  E_\nu(\bp)  < E_\nu(\bzero) \quad \text{for all } \bp\neq \bzero.
\end{equation*}
\end{prop}
\begin{rem}
 When $\nu>0$, the massive Hamiltonian $H_\nu(\bzero)$ has a ground state
(Lemma \ref{L mgs}). In the massless case $\nu=0$, $H(\bzero)$ has 
a ground state under suitable conditions(see Theorems \ref{T1}, \ref{T2} and \ref{T3}.)
\end{rem}

\begin{proof}[Proof of Proposition \ref{P1.75}]
 We assume the equality $E_\nu(\bp)=E_\nu(\bzero)$ for a nonzero vector 
$\bp\in\BR^3\setminus\{\bzero\}$. 
Let $\Phi_\nu(\bzero)$ be a normalized ground state of $H_\nu(\bzero)$.
For $t=1,-1$, we have
\begin{equation*}
 E_\nu(\bp)=E_\nu(t\bp) \leq \inner{\Phi_\nu(\bzero)}{H_\nu(t\bp)\Phi_\nu(\bzero)}
            =   t\inner{\Phi_\nu(\bzero)}{\balpha\cdot\bp\Phi_\nu(\bzero)}+ E_\nu(\bzero).
\end{equation*}
Therefore $\inner{\Phi_\nu(\bzero)}{\balpha\cdot\bp \Phi_\nu(\bzero)}=0$, and hence
$\inner{\Phi_\nu(\bzero)}{H_\nu(\bp)\Phi_\nu(\bzero)}=E_\nu(\bzero)=E_\nu(\bp)$, 
which implies $\norm{(H_\nu(\bp)-E_\nu(\bp))^{1/2}\Phi_\nu(\mathbf{0})}=0$,
and therefore, $\Phi_\nu(\bzero)$ is a ground state of $H_\nu(\bp)$.
Thus $\balpha\cdot \bp \Phi_\nu(\bzero)=0$, and we get a contradiction
$|\bp|^2\Phi_\nu(\bzero)=0$.
\end{proof}
If the cutoff function $\hat\rho$ is spherically symmetric,
the spectral properties of $\overline{H_\nu(\bp)}$ is independent 
of the direction of $\bp$. 
The first part of the following proposition immediately follows from Proposition \ref{P1.6},
and thus, the last part from Proposition \ref{P1.3}.
\begin{prop}[Spherical symmetry in the total momentum]{\label{P1.8}}
 Assume that $|\hat\rho(\bk)|$ is a spherically symmetric function. 
Then $\overline{H_\nu(\bp)}$ is unitarily equivalent to 
$\overline{H_\nu(\bp')}$ for all $\bp'\in\BR^3$ with $|\bp|=|\bp'|$.
In particular $E_\nu(\bp)$ is spherically symmetric with respect to $\bp$, and 
$E_\nu(\bp)\geq E_\nu(\bp')$ if $|\bp|\leq |\bp'|$.
\end{prop}
\begin{prop}[Massless limit]{\label{P1.9}}
 $E_\nu(\bp)$ is monotonously non-decreasing in $\nu\geq 0$ and 
\begin{equation*}
 \lim_{\nu\to +0} E_\nu(\bp) = E_0(\bp).
\end{equation*}
\end{prop}
\begin{proof}
 Let $\nu\geq \nu'\geq 0$. Then we have $H_\nu(\bp) \geq H_{\nu'}(\bp)$ in the 
sense of quadratic form on $\cD:=\dom(H_f)\cap \dom(N_f)$. 
Therefore $\nu\mapsto E_\nu(\bp)$ is non-decreasing: 
$E_\nu(\bp)\geq E_{\nu'}(\bp)$.
It is easy to see that for all $\Psi\in \cD $,
$H_\nu(\bp)\Psi\to H(\bp)\Psi$ as $\nu\to 0$.
Since $\cD $ is a common core for all $H_\nu(\bp)$, 
$H_\nu(\bp)\to H(\bp)$ in the strong resolvent sense
(see \cite[Theorem VIII. 25]{Reed-Simon-I}).
Using a fact about a strongly convergent 
operators\cite[Theorem VIII. 24]{Reed-Simon-I}, we obtain that
$E_\nu(\bp)\to E(\bp)$ as $\nu\to+0$.
\end{proof}

By Proposition \ref{P1.4}, the following inequality holds:
\begin{equation*}
   0\leq E_\nu(\bp-\bk) -E_\nu(\bp) + |\bk|, \quad \bp,\bk\in\BR^3.
\end{equation*}
The function $\bk\to E_\nu(\bp-\bk)-E_\nu(\bp)+|\bk|$ plays the role of a
dispersion relation in the low-energy Dirac polaron.
\begin{thm}{\label{T LBOD}}
Let $\nu\geq 0$.
Assume that $\hat\rho$ is spherically symmetric.
Suppose that $E_\nu(\bp,m)< E_\nu(\bp,0)$.
Then, for $\bp\neq \bzero$, the following estimate holds:
\begin{equation*}
  E_\nu(\bp-\bk,m) - E_\nu(\bp,m) +|\bk| \geq 
  \begin{cases}
    |\bk|  \quad     &\text{if  }~~ |\bp-\bk|\leq |\bp|, \\
    (1-b_\nu(\bp))|\bk| \quad &\text{if  }~~ |\bp|\leq |\bp-\bk| \leq 2|\bp|, \\
    (1-b_\nu(\bp))|\bp| \quad &\text{if  }~~  2|\bp| \leq |\bp-\bk|,
  \end{cases}
\end{equation*}
where 
\begin{equation*}
  b_\nu(\bp) := \frac{E_\nu(\bp,m)-E_\nu(2\bp,m)}{|\bp|} < 1.
\end{equation*}
In the case $\bp=\bzero$, for all constant $P>0$ the following estimate holds:
\begin{align} 
 E_\nu(\bk,m)-E_\nu(\bzero,m)+|\bk| \geq 
 \begin{cases}
  \frac{a_\nu(P)}{P} |\bk|,  &\text{if}~~ |\bk|\leq P  \\
   a_\nu(P)        ,         &\text{if}~~ |\bk| > P,   
 \end{cases} \label{e p6}
\end{align}
where
\begin{equation*}
 a_\nu(P) := (E_\nu(\bk,m)-E_\nu(\bzero,m)+|\bk|)\Big|_{|\bk|=P}
\end{equation*}
is a strictly positive constant.
\end{thm}
\begin{rem}
The idea of the proof of Theorem \ref{T LBOD} was developed in \cite{Loss-Miyao-Spohn:2007}.
\end{rem}

\begin{proof}[Proof of Theorem \ref{T LBOD}]
Before proving Theorem \ref{T LBOD}, we prove the next lemma:
\begin{lem}{\label{L 6.1}}
Let $\nu\geq 0$. Assume that $E_\nu(\bp,m)<E_\nu(\bp,0)$. Then
  \begin{equation}
    E_\nu(\bp-\bk,m) -E_\nu(\bp,m) +|\bk|>0, \quad \bk\in\BR^3\backslash\{\bzero\}.
     \label{c12ineq}
  \end{equation}
\end{lem}
\begin{proof}
  First we prove \eqref{c12ineq} for positive $\nu>0$.
We fix $m\neq 0$ and $\bp\in\BR^3$. Suppose that 
\begin{equation}
  E_\nu(\bp-\bk)-E_\nu(\bp) +|\bk|=0,  \label{51}
\end{equation}
for some $\bk\in\BR^3\backslash\{\bzero\}$. Let $\Phi_\nu(\bp-\bk)$ be a normalized
ground state of $H_\nu(\bp-\bk)$(see Lemma \ref{L mgs}). Then
\begin{align*}
 E_\nu(\bp-\bk)
 &= \inner{\Phi_\nu(\bp-\bk)}{\overline{H_\nu(\bp-\bk)}\Phi_\nu(\bp-\bk)} \\
 &=
 \inner{\Phi_\nu(\bp-\bk)}{\overline{H_\nu(\bp)}\Phi_\nu(\bp-\bk)}
 -
 \inner{\Phi_\nu(\bp-\bk)}{\balpha\cdot\bk\Phi_\nu(\bp-\bk)} \\
 &\geq 
 E_\nu(\bp) - |\bk|.
\end{align*}
Hence, by assumption (\ref{51}) we have
 $\inner{\Phi_\nu(\bp-\bk)}{\overline{H_\nu(\bp)}\Phi_\nu(\bp-\bk)}=E_\nu(\bp)$ and
$\inner{\Phi_\nu(\bp-\bk)}{\balpha\cdot\bk\Phi_\nu(\bp-\bk)}=|\bk|$, which implies
that $\Phi_\nu(\bp-\bk)$ is a ground state of both $\overline{H_\nu(\bp)}$ and 
$-\balpha\cdot\bk$.
Since $\bk \neq \bzero$, we have 
$\inner{\Phi_\nu(\bp-\bk)}{\beta\Phi_\nu(\bp-\bk)}=0$,
 because $\balpha\cdot\bk\beta=-\beta\balpha\cdot\bk$.
In what follows, to emphasize $m$-dependence,
 we write $H_\nu(\bp-\bk,m)$ and $\Phi_\nu(\bp-\bk,m)$ for $H_\nu(\bp-\bk)$ and  $\Phi_\nu(\bp-\bk)$, respectively. 
By using the above facts, we have
\begin{equation*}
 E_\nu(\bp,m) = 
 \inner{\Phi_\nu(\bp-\bk,m)}{\overline{H_\nu(\bp,0)}\Phi_\nu(\bp-\bk,m)}
 \geq 
 E_\nu(\bp,0),
\end{equation*}
which contradicts the inequality $E_\nu(\bp,m)<E_\nu(\bp,0)$.
Next, we prove the case $\nu=0$. 
Suppose that 
there exist a vector $\bk\in\BR^3\backslash\{\bzero\}$ such that
$E(\bp-\bk,m)-E(\bp,m)+|\bk|=0$ holds.
It is not difficult to see that 
\begin{align*}
 &\lim_{\nu\to+0}\inner{\Phi_\nu(\bp-\bk,m)}{\overline{H(\bp-\bk,m)}\Phi_\nu(\bp-\bk,m)}=E(\bp-\bk,m).
\end{align*}
By these equations, we have 
\begin{align}
 &\lim_{\nu\to+0}
 \inner{\Phi_\nu(\bp-\bk,m)}{\balpha\cdot\bk\Phi_\nu(\bp-\bk,m)}
  =  |\bk|,    \label{eq 17}\\
 &\lim_{\nu\to+0}
 \inner{\Phi_\nu(\bp-\bk,m)}{\overline{H(\bp,m)}\Phi_\nu(\bp-\bk,m)}
 = E(\bp,m).   \label{eq 18}
\end{align}
Equation (\ref{eq 17}) implies that 
\begin{align*}
  \lim_{\nu\to+0} (|\bk|-\balpha\cdot\bk)\Phi_\nu(\bp-\bk,m)=0.
\end{align*}
Therefore 
$\lim_{\nu\to+0}\inner{\Phi_\nu(\bp-\bk,m)}{\beta\Phi_\nu(\bp-\bk,m)}=0$.
This fact and equation (\ref{eq 18}) imply $E(\bp,m)=E(\bp,0)$,
which contradicts $E(\bp,m)<E(\bp,0)$.
\end{proof}
We fix a vector $\bp$ such that $E_\nu(\bp,m)<E_\nu(\bp,0)$.
Since $\hat\rho$ is spherically symmetric, by Proposition \ref{P1.8}, the function 
\begin{equation*}
  G_\nu(|\bk|) := E_\nu(\bzero) - E_\nu(\bk), \quad \bk\in\BR^3,
\end{equation*}
is monotonously non-decreasing, convex with respect to $|\bk|$, and
the following inequality holds 
\begin{equation}
 0 \leq G_\nu(|\bk|) \leq |\bk|, \quad \bk\in\BR^3. \label{G1}
\end{equation}
Since $G_\nu(s)$ is convex, $G_\nu(s)$ has a right derivative ${G_\nu^+}'(s)$:
\begin{equation*}
  {G_\nu^+}'(s) := \lim_{h\to+0} [G_\nu(s+h)-G_\nu(s)]/h.
\end{equation*}
First we show that 
\begin{equation}
  {G_\nu^+}'(s) <1, \quad 0\leq s \leq |\bp|.  \label{G}
\end{equation}
  Since $G_\nu(s)$ is convex and $0\leq G_\nu(s)\leq s$, ${G_\nu^+}'(s)$ is a 
 monotonously non-decreasing function of $s$.
If ${G_\nu^+}'(s_0)>1$ for a constant $s_0\geq 0$, 
then ${G_\nu^+}'(s)>1$ for all $s\geq s_0$ and 
  \begin{equation*}
    G_\nu(s) = \int_{s_0}^s {G_\nu^+}'(t) d t +\int_0^{s_0}{G_\nu^+}'(t)  d t
           \geq (s-s_0) {G_\nu^+}'(s_0) + \int_0^{s_0}{G_\nu^+}'(t) d t,
  \end{equation*}
holds for all $s>s_0$.
It contradicts (\ref{G1}). Thus, ${G_\nu^+}'(s)\leq 1$ for all $s\geq 0$.
Let $s_1\geq 0$ be a point such that ${G_\nu^+}'(s_1)=1$ and ${G_\nu^+}'(s_1-\ep)<1$
for all $0< \ep\leq s_1$.
If $|\bp|<s_1$, (\ref{G}) is trivial. 
Thus we consider the case $|\bp|\geq s_1$.
Note that ${G_\nu^+}'(s)=1$ for all $s\geq s_1$.
Hence $G_\nu(s)$ is a linear function of $s$ if $s\geq s_1$:
\begin{equation*}
  G_\nu(s) = s + C, \quad s\geq s_1,
\end{equation*}
where $C$ is a negative constant.
By this equality, we have that 
\begin{equation*}
  E_\nu(\bp-\bk) -E_\nu(\bp) + |\bk| = -|\bp-\bk| +|\bp| +|\bk|,
\end{equation*}
for all $\bp$ and $\bk$ such that $|\bp-\bk|\geq s_1$ and $|\bp|\geq s_1$.
We choose $\bk=-C\bp$ for a constant $C>s_1/|\bp|$. Then
\begin{equation*}
  E_\nu(\bp-\bk) - E_\nu(\bp) +|\bk| = 0.
\end{equation*}
It contradicts Lemma \ref{L 6.1}. Therefore
${G_\nu^+}'(s)<1$ holds for all $0\leq s \leq |\bp|$. 

Next, by using this inequality, we prove Theorem \ref{T LBOD}.
By (\ref{G}) and convexity of $G_\nu$, it holds that
\begin{equation*}
  c_\nu(\bp):= \frac{G_\nu(|\bp|)}{|\bp|} \leq b_\nu(\bp) <1.
\end{equation*}
  We define a set of functions:
\begin{align*}
 \mathcal{C} := \{ J:\BR_+\to \BR_+|& J \text{ is convex}, ~~
                                      0\leq J(s) \leq s, ~ (s\geq 0)\\
                                    &J(|\bp|)=G_\nu(|\bp|), J(2|\bp|)=G_\nu(2|\bp|)
                \}    
\end{align*}
Then we have
\begin{align}
  E_\nu(\bp-\bk) - E_\nu(\bp) + |\bk| &= |\bk| +G_\nu(\bp) - G_\nu(\bp-\bk) \notag \\
                              &\geq |\bk| + G_\nu(\bp) -\sup_{J\in\mathcal{C}}J(\bp-\bk). 
  \label{kgj}
\end{align}
The maximal function in $\mathcal{C}$ is given by the following linear interpolation:
\begin{equation*}
  J_{\mathrm{max}}(s) := 
  \begin{cases}
    c_\nu(\bp) s      \quad               &\text{if} ~~ s\leq |\bp|, \\
    b_\nu(\bp)(s-|\bp|) + G_\nu(|\bp|) \quad&\text{if} ~~ |\bp|\leq s\leq 2|\bp|, \\
    s-2|\bp|+G_\nu(2|\bp|)           \quad &\text{if} ~~ 2|\bp| \leq |\bp-\bk|.
  \end{cases}
\end{equation*}
Hence 
\begin{align*}
  (\ref{kgj}) &\geq  |\bk|+G_\nu(|\bp|) -
  \begin{cases}
   c_\nu(\bp) |\bp-\bk| \quad &\text{if} ~~ |\bp-\bk| \leq |\bp|, \\
   b_\nu(\bp)(|\bp-\bk|-|\bp|) +G_\nu(|\bp|) &\text{if} ~~ |\bp|\leq |\bp-\bk| \leq 2|\bp|,\\
   |\bp-\bk| -2|\bp| +G_\nu(2|\bp|)        &\text{if} ~~ 2|\bp| \leq |\bp-\bk|.
  \end{cases} \\
 &= 
 \begin{cases}
  |\bk| + c_\nu(\bp)(|\bp|-|\bp-\bk|) \quad &\text{if} ~~ |\bp-\bk| \leq |\bp|, \\
  |\bk| - b_\nu(\bp)(|\bp-\bk|-|\bp|) &\text{if} ~~ |\bp| \leq |\bp-\bk| \leq 2|\bp|, \\
  |\bk|-|\bp-\bk|+(2-b_\nu(\bp))|\bp|  &\text{if} ~~ 2|\bp|\leq |\bp-\bk|.
 \end{cases}
\end{align*}
Using the triangle inequality, one can obtain the desired estimate.
Finally we prove (\ref{e p6}). 
Since ${G_\nu^+}'(0)<1$ and $G_\nu$ is convex,
the constant $a_\nu(P)$ is strictly positive for all $P>0$.
It is easy to see that 
\begin{equation*}
  {G_\nu^+}'(s) \leq \frac{G_\nu(P)}{P} = \frac{-a_\nu(P)+P}{P}, \quad s\leq P.
\end{equation*}
Hence
\begin{align*}
  E_\nu(\bk)-E_\nu(\bzero)+|\bk| &= |\bk|-G_\nu(|\bk|) 
                       = \int_0^{|\bk|}(1-{G_\nu^+}'(s))  d s \\
    & \geq
    \begin{cases}  \displaystyle
      \int_0^{|\bk|} \Big(1-\frac{G_\nu(P)}{P}\Big) d s  &~~\text{if}~~ |\bk|\leq P.\\
       \displaystyle
      \int_0^{P} \Big(1-\frac{G_\nu(P)}{P}\Big) d s
       + \int_{P}^{|\bk|} (1-{G_\nu^+}'(s)) d s          &~~\text{if}~~ |\bk|> P.
    \end{cases} \\
    & \geq 
    \begin{cases}
      (a_\nu(P)/P)|\bk|,     &~~\text{if}~~ |\bk|\leq P. \\
      a_\nu(P)         ,     &~~\text{if}~~ |\bk|>P.
    \end{cases}
\end{align*}
This completes the proof.
\end{proof}


\section*{Acknowledgments}
I would like to thank A. Arai for his advice, discussions and encouragement.
I am grateful to F. Hiroshima and T. Miyao for their advices.

\bibliographystyle{amsplain}
\bibliography{sasaki-biblio}

\providecommand{\bysame}{\leavevmode\hbox to3em{\hrulefill}\thinspace}
\providecommand{\MR}{\relax\ifhmode\unskip\space\fi MR }
\providecommand{\MRhref}[2]{%
  \href{http://www.ams.org/mathscinet-getitem?mr=#1}{#2}
}
\providecommand{\href}[2]{#2}
\begin{thebibliography}{10}

\bibitem{Arai:1999}
Asao Arai, \emph{Fundamental properties of the hamiltonian of a dirac particle
  coupled to the quantized radiation field}, Hokkaido Univ.Preprint Series in
  Math (1999), no.~447.

\bibitem{Arai:2000}
\bysame, \emph{A particle-field {H}amiltonian in relativistic quantum
  electrodynamics}, J. Math. Phys. \textbf{41} (2000), no.~7, 4271--4283.

\bibitem{Arai:2003}
\bysame, \emph{Non-relativistic limit of a {D}irac-{M}axwell operator in
  relativistic quantum electrodynamics}, Rev. Math. Phys. \textbf{15} (2003),
  no.~3, 245--270.

\bibitem{Arai:2006}
\bysame, \emph{Non-relativistic limit of a {D}irac polaron in relativistic
  quantum electrodynamics}, Lett. Math. Phys. \textbf{77} (2006), no.~3,
  283--290.

\bibitem{Bjorken-Drell:1964}
James~D. Bjorken and Sidney~D. Drell, \emph{Relativistic quantum mechanics},
  McGraw-Hill Book Co., New York, 1964.

\bibitem{Chen:2001}
T.~{Chen}, \emph{{Operator-theoretic infrared renormalization and construction
  of dressed 1-particle states in non-relativistic QED}}, ArXiv Mathematical
  Physics e-prints (2001).

\bibitem{Gerard:2000}
C.~G{\'e}rard, \emph{On the existence of ground states for massless
  {P}auli-{F}ierz {H}amiltonians}, Ann. Henri Poincar\'e \textbf{1} (2000),
  no.~3, 443--459.

\bibitem{Griesemer-Lieb-Loss:2001}
M.~Griesemer, E.~H. Lieb, and Loss. M., \emph{Ground states in non-relativistic
  quantum electrodynamics}, Invent Math \textbf{145} (2001), no.~1, 557--595.

\bibitem{Heitler:1954}
Walter Heitler, \emph{The quantum theory of radiation}, Oxford University
  Press, 1954.

\bibitem{Hiroshima:2007}
Fumio Hiroshima, \emph{Fiber {H}amiltonians in non-relativistic quantum
  electrodynamics}, J. Funct. Anal. \textbf{252} (2007), no.~1, 314--355.

\bibitem{lieb-loss-analysis}
E.~H. Lieb, , and M.~Loss, \emph{Analysis}, Graduate Studies in Mathematics
  Series, Amer Mathematical Society, 2001.

\bibitem{Loss-Miyao-Spohn:2007}
Michael Loss, Tadahiro Miyao, and Herbert Spohn, \emph{Lowest energy states in
  nonrelativistic {QED}: atoms and ions in motion}, J. Funct. Anal.
  \textbf{243} (2007), no.~2, 353--393.

\bibitem{Reed-Simon-I}
Michael Reed and Barry Simon, \emph{Methods of modern mathematical physics.
  {I}. {F}unctional analysis}, Academic Press, New York, 1972.

\bibitem{Reed-Simon-II}
\bysame, \emph{Methods of modern mathematical physics. {II}. {F}ourier
  analysis, self-adjointness}, Academic Press, New York, 1975.

\bibitem{Sasaki:2005}
Itaru Sasaki, \emph{Ground state energy of the polaron in the relativistic
  quantum electrodynamnics}, J. Math. Phys. \textbf{46} (2005), no.~10, 102307,
  6.

\bibitem{Sasaki:2006}
\bysame, \emph{Ground state of the polaron in the relativistic quantum
  electrodynamics}, RIMS Kokyuroku \textbf{1510} (2006), no.~10, 87--103.

\bibitem{Spohn:2004}
Herbert Spohn, \emph{Dynamics of charged particles and their radiation field},
  Cambridge University Press, Cambridge, 2004.

\bibitem{Thaller:1992}
Bernd Thaller, \emph{The {D}irac equation}, Texts and Monographs in Physics,
  Springer-Verlag, Berlin, 1992.

\end{thebibliography}

\end{document}